\documentclass[aps, prl, twocolumn, superscriptaddress, amsmath,  tightenlines, longbibliography]{revtex4-1}

\usepackage{dcolumn}
\usepackage{graphicx}
\usepackage{mathrsfs}
\usepackage{subfigure}
\usepackage{booktabs}
\usepackage{amsmath}
\usepackage{physics}
\usepackage{dsfont}
\usepackage{amstext}
\usepackage{amssymb}
\usepackage{amsbsy}
\usepackage{bbm}
\usepackage{amsthm}
\usepackage{graphicx}
\usepackage{color}
\usepackage[colorlinks,citecolor=blue]{hyperref}

\usepackage{url}
\usepackage[colorlinks]{hyperref}
\hypersetup{%
	plainpages=true,
	breaklinks=true,       
	hypertexnames=false,  
	pageanchor=true,
	colorlinks=true,
	linkcolor={blue},
	citecolor={red},
	urlcolor={blue},
	anchorcolor={black}
}

 \makeatletter

\newcommand{\Rmnum}[1]{\expandafter\@slowromancap\romannumeral #1@}
\makeatother

\begin{document}

\title{Second-Order Topological Phases in Non-Hermitian Systems}
\author{Tao Liu}
\email{tao.liu@riken.jp}
\affiliation{Theoretical Quantum Physics Laboratory, RIKEN Cluster for Pioneering Research, Wako-shi, Saitama 351-0198, Japan}

\author{Yu-Ran Zhang}
\affiliation{Theoretical Quantum Physics Laboratory, RIKEN Cluster for Pioneering Research, Wako-shi, Saitama 351-0198, Japan}
\affiliation{Beijing Computational Science Research Center, Beijing 100193, China}

\author{Qing Ai}
\affiliation{Theoretical Quantum Physics Laboratory, RIKEN Cluster for Pioneering Research, Wako-shi, Saitama 351-0198, Japan}
\affiliation{Department of Physics, Applied Optics Beijing Area Major Laboratory, Beijing Normal University, Beijing 100875, China}

\author{Zongping Gong}
\email{gong@cat.phys.s.u-tokyo.ac.jp}
\affiliation{Department of Physics, University of Tokyo, 7-3-1 Hongo, Bunkyo-ku, Tokyo 113-0033, Japan}

\author{Kohei Kawabata}
\email{kawabata@cat.phys.s.u-tokyo.ac.jp}
\affiliation{Department of Physics, University of Tokyo, 7-3-1 Hongo, Bunkyo-ku, Tokyo 113-0033, Japan}

\author{Masahito Ueda}
\email{ueda@phys.s.u-tokyo.ac.jp}
\affiliation{Department of Physics, University of Tokyo, 7-3-1 Hongo, Bunkyo-ku, Tokyo 113-0033, Japan}
\affiliation{RIKEN Center for Emergent Matter Science (CEMS), Wako, Saitama 351-0198, Japan}

\author{Franco Nori}
\email{fnori@riken.jp}
\affiliation{Theoretical Quantum Physics Laboratory, RIKEN Cluster for Pioneering Research, Wako-shi, Saitama 351-0198, Japan}
\affiliation{Department of Physics, University of Michigan, Ann Arbor, Michigan 48109-1040, USA}



\begin{abstract}
A $d$-dimensional second-order topological insulator (SOTI) can host topologically protected  $(d - 2)$-dimensional gapless boundary modes. Here we show that a 2D non-Hermitian SOTI can host zero-energy modes at its corners. In contrast to the Hermitian case, these zero-energy modes can be localized only at one corner. A 3D non-Hermitian SOTI is shown to support second-order boundary modes, which are localized not along hinges but anomalously at a corner. The usual bulk-corner (hinge) correspondence in the second-order 2D (3D) non-Hermitian system breaks down. The winding number (Chern number) based on complex wavevectors is used to characterize the second-order topological phases in 2D (3D). A possible experimental situation with ultracold atoms is also discussed. Our work lays the cornerstone for exploring higher-order topological phenomena in non-Hermitian systems.
\end{abstract}

\maketitle


\textit{Introduction}.---Recent years have witnessed a surge of theoretical and experimental interest in studying topological phases \cite{RevModPhys.82.3045, RevModPhys.83.1057,  RevModPhys.88.035005} in insulators \cite{PhysRevLett.61.2015, PhysRevLett.95.146802, Bernevig1757, zhangTI2007, PhysRevB.75.121306, Zhang2009}, superconductors \cite{RPPJason2012, Beenakker2013, RPPMasatoshi2017}, ultracold atoms \cite{Aidelsburger2014, Jotzu2014, Lohse2015, Nakajima2016, Goldman2016, arXiv:1803.00249} and classical waves \cite{Khanikaev2012, Horiuchi2013, Sebastian2015, Khanikaev2017}. These topologically nontrivial phases are characterized by the topological index of gapped bulk energy bands and exhibit gapless states on their boundaries. Such gapless boundary states cannot be gapped out by local perturbations that preserve both bulk gap and symmetry. 

Topological phases have widely been studied in closed systems, which are described by Hermitian Hamiltonians featuring real eigenenergies and orthogonal eigenstates. Recently, there has been a great deal of effort in exploring topological invariants of open systems governed by non-Hermitian operators \cite{PhysRevLett.80.5243, Bender2007}. Non-Hermitian Hamiltonians can find applications in a wide range of systems including optical and mechanical structures subjected to gain and loss \cite{PhysRevLett.100.103904, PhysRevLett.106.093902, Regensburger2012, PhysRevLett.113.053604, Hodaei975, Peng2014, Feng972, Peng328, Jing2015, PhysRevLett.117.110802, PhysRevLett.119.190401, Jing2017, PhysRevApplied.8.044020, Ashida2017, El-Ganainy2018, Zhang2018}, and solid-state systems with finite quasiparticle lifetimes \cite{arXiv:1802.03023, PhysRevB.97.041203, arXiv:1802.00443, arXiv:1708.05841,PhysRevB.98.035141}. In particular, topological phases of non-Hermitian Hamiltonians have recently been investigated in these systems \cite{PhysRevLett.102.065703, PhysRevB.84.205128, PhysRevB.98.035141, PhysRevLett.115.200402, Weimann2016, PhysRevLett.116.133903, PhysRevLett.118.045701, PhysRevLett.118.040401, PhysRevLett.120.146402, PhysRevB.97.121401, Harari2018, Bandreseaar4005, Zhoueaap9859, Xiong2018, Pan2018, arXiv:1802.07964, arXiv:1806.06566, arXiv:1804.04676, arXiv:1805.06492, arXiv:1801.00499, ShunyuYao2018, YaoarXiv:1804.04672, KoheiarXiv:1805.09632, arXiv:1802.00443, arXiv:1708.05841, arXiv:1808.06162, arXiv:1809.02125,arXiv:1901.00346}. The most prominent feature of non-Hermitian Hamiltonians is the existence of exceptional points (EPs), where more than one eigenstate coalesces \cite{Berry2004, Bender2007,  Heiss2012}. This coalescence of eigenstates at EPs makes the corresponding eigenspace no longer complete, and the non-Hermitian Hamiltonian becomes non-diagonalizable. These unique features of EPs can lead to rich topological features in non-Hermitian topological systems with no counterpart in Hermitian cases such as Weyl exceptional rings \cite{PhysRevLett.118.045701}, bulk Fermi arcs, and half-integer topological charges \cite{Zhoueaap9859}. Furthermore, the interplay between non-Hermiticity and topology can lead to the breakdown of the usual bulk-boundary correspondence \cite{PhysRevLett.116.133903, PhysRevLett.118.040401, KoheiarXiv:1805.09632, Xiong2018, arXiv:1805.06492, ShunyuYao2018,YaoarXiv:1804.04672} due to the non-Bloch-wave behavior of open-boundary eigenstates, where the conventional Bloch wavefunctions do not precisely describe topological phase transitions under the open boundary conditions. The non-Bloch winding (Chern) number defined via complex wavevectors in 1D (2D) has recently been introduced to fill this gap \cite{ShunyuYao2018,YaoarXiv:1804.04672}.

More recently, the concept of topological insulators (TIs) has been generalized to second-order \cite{PhysRevLett.110.046404, Benalcazar61, PhysRevB.96.245115, PhysRevLett.119.246401, PhysRevB.97.241405, Peterson2018, Serra-Garcia2018, PhysRevLett.119.246402, Imhof2018, TitusSciAdv2018,  PhysRevB.97.155305, arXiv:1802.02585,  PRBXYZhu2018, Noh2018, arXiv:1801.10053, arXiv:1803.08545, arXiv:1804.04711, arXiv:1806.07002, arXiv:1804.02794} and third-order \cite{Benalcazar61, PhysRevLett.120.026801, arXiv:1801.10050} TIs  in Hermitian systems. In contrast to conventional first-order TIs, a $d$-dimensional second-order topological insulator (SOTI) only hosts topologically protected  $(d - 2)$-dimensional gapless boundary states. For example, a 2D SOTI has zero-energy states localized at its corners, and a 3D SOTI hosts 1D gapless modes along its hinges. Therefore, the conventional bulk-boundary correspondence is no longer applicable to SOTIs. Up to now, studies of the second-order and third-order topological phases have been restricted to Hermitian systems. We now ask: is it possible for a non-Hermitian system to exhibit second-order topological phases? If yes, how can we define a topological invariant to characterize them?

In this Letter, we investigate 2D and 3D SOTIs described by non-Hermitian Hamiltonians. Even though the bulk bands are first-order topologically trivial insulators, there are degenerate second-order bound states. In contrast to the Hermitian case, these zero-energy states in 2D are localized only at one corner protected by mirror-rotation symmetry and sublattice symmetry. Moreover, the second-order boundary modes in 3D are localized not along the hinges but anomalously at a corner. The winding number (Chern number) characterizes its second-order topological phase in 2D (3D), where the non-Bloch-wave behavior of open-boundary eigenstates is included due to the breakdown of the usual bulk-corner (hinge) correspondence in second-order non-Hermitian systems. The proposed non-Hermitian model can experimentally be realized in ultracold atoms.  

\textit{\textrm{2D} SOTI}.---We consider a 2D non-Hermitian Hamiltonian $H_{2D}$ that respects both two-fold mirror-rotation symmetry $ {\cal M}_{xy} $ and sublattice symmetry ${\cal S}$
\begin{equation}
{\cal M}_{xy} H_{2D} \left( k_{x}, k_{y} \right) {\cal M}_{xy}^{-1}
= H_{2D} \left( k_{y}, k_{x}\right),
\end{equation}
\begin{equation}
{\cal S} H \left( k_{x}, k_{y} \right) {\cal S}^{-1} = - H \left( k_{x}, k_{y} \right),
\end{equation}
and $\left[{\cal S}, ~{\cal M}_{xy}\right] = 0$. Note that the Hermitian counterpart with the same symmetries was investigated in Ref.~\cite{Imhof2018}. Due to the mirror-rotation symmetry in Eq.~(1), we can express the Hamiltonian $H_{2D}$ on the high-symmetry line $k_{x} = k_{y}$ as
\begin{equation}
U^{-1} H_{2D} \left(k, k \right) U = \left( \begin{array}{@{\,}cc@{\,}} 
H_{+} \left( k \right) & 0 \\
0 & H_{-} \left( k \right) \\ 
\end{array} \right),
\end{equation}
where $U$ is a unitary operator, and $H_{\pm}(k)$ acts on the mirror-rotation subspace. Since $H_{\pm} \left( k \right)$ respects sublattice symmetry ${\cal S}'$ defined in each mirror-rotation subspace [note that $\mathcal{S}$ in Eq.~(2) is defined in the entire lattice space], we can define the winding number as follows:
\begin{equation}
w_{\pm} := \oint_{\rm BZ} \frac{dk}{4\pi i }\,{\rm Tr} \left[ {{\cal S'}} H_{\pm}^{-1} \left( k \right) \frac{dH_{\pm} \left( k \right)}{dk} \right].
\end{equation}
The topological index that characterizes the second-order topological phases in 2D is given by
\begin{align}
w := w_{+} - w_{-}.
\end{align} 

We investigate a concrete model of a 2D SOTI on a square lattice, where each unit cell contains four orbitals and asymmetric particle hopping within each unit cell is introduced, as shown in Fig.~1(a). The Bloch Hamiltonian is written as
\begin{align}
H_{\textrm{2D}} & = \left[t + \lambda \cos(k_x)\right] \tau_x - \left[\lambda \sin(k_x) ~ + i \gamma \right] \tau_y \sigma_z     \nonumber \\
&~~~~  + \left[t + \lambda \cos(k_y)\right] \tau_y \sigma_y + \left[\lambda \sin(k_y) + i \gamma\right] \tau_y \sigma_x, 
\end{align}
where we have set the lattice constant $a_0 = 1$, $\lambda$ is a real-valued  inter-cell hopping amplitude, $t ~\pm~ \gamma$ denote real-valued asymmetric intra-cell hopping amplitudes, and $\sigma_i$ and $\tau_i$ ($i = x, y, z$) are Pauli matrices for the degrees of freedom within a unit cell. The Hamiltonian $H_{\textrm{2D}}$ can be implemented experimentally using ultracold atoms in optical lattices with engineered dissipation [see Fig.~1(b) and  Sec.~$\textrm{\Rmnum{8}}$ in the supplemental material \cite{SupplMaterialsNonhermitian2018} for details]. The Hermitian part of $H_{\textrm{2D}}(\mathbf{k})$ preserves mirror and four-fold rotational symmetries with $\mathcal{M}_x = \tau_x \sigma_z$, $\mathcal{M}_y = \tau_x \sigma_x$, and $C_4 = \left[(\tau_x - i \tau_y) \sigma_0 - (\tau_x + i \tau_y) (i\sigma_y) \right]/2$. While they are broken by asymmetric hopping,  $H_{\textrm{2D}}$ stays invariant under sublattice symmetry $\mathcal{S} = \tau_z$ and mirror-rotation symmetry $\mathcal{M}_{xy} = C_4 \mathcal{M}_y$, and $\left[{\cal S}, ~\mathcal{M}_{xy}\right] = 0$.

\begin{figure}[!tb]
	\centering
	\includegraphics[width=8.4cm]{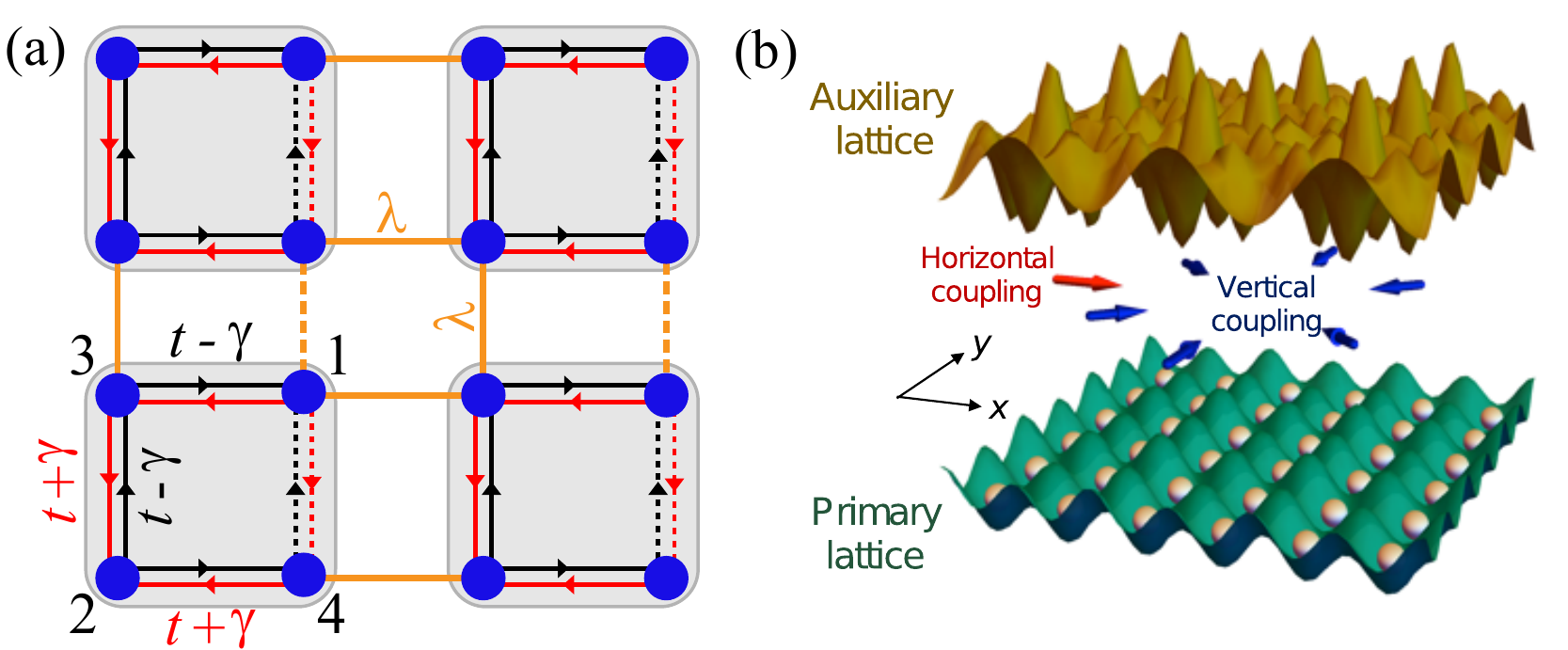}
	\caption{Non-Hermitian SOTI in 2D. (a) Tight-binding representation of the model [Eq.~(6)] on a square lattice. Each unit cell contains four orbitals (blue solid circles). The orange lines denote inter-cell coupling, and the red and black lines with arrows represent asymmetric intra-cell hopping. The dashed lines indicate hopping terms with a negative sign, accounting for a flux of $\pi$ piercing each plaquette. (b) Schematic illustration of a proposed experimental setup using ultracold atoms \cite{SupplMaterialsNonhermitian2018}. The primary lattice together with a pair of Raman lasers gives rise to a Hermitian SOTI, where the Raman lasers are used for inducing effective particle hopping. The asymmetric hopping amplitudes are introduced via  coherent coupling to a dissipative auxiliary lattice.}\label{fig1}
\end{figure}
\begin{figure}[!tb]
	\centering
	\includegraphics[width=8.4cm]{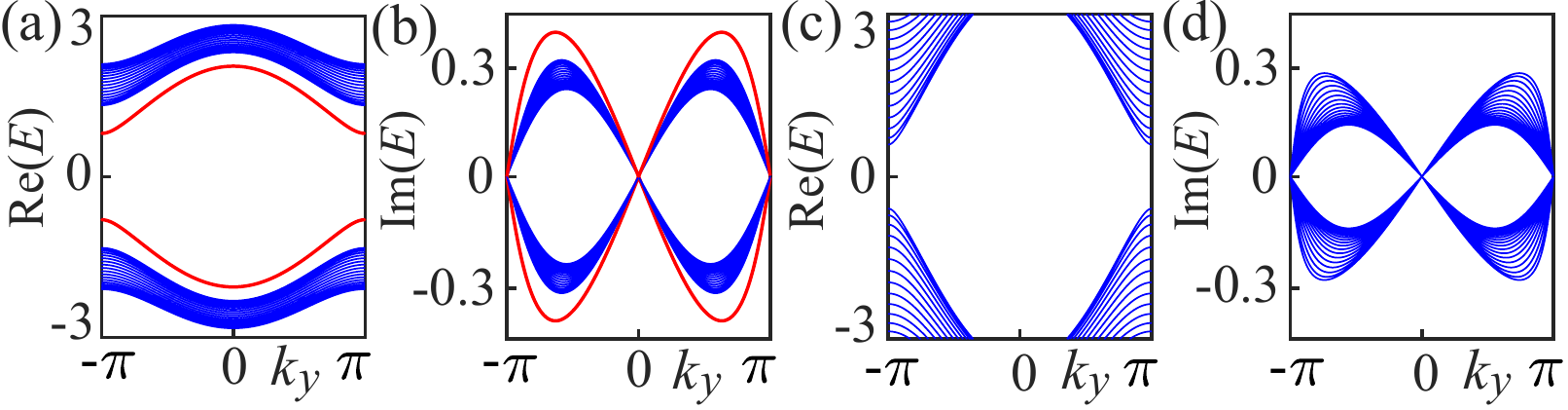}
	\caption{Complex energy spectra of the non-Hermitian SOTI described by Eq.~(6) with open boundaries along the $x$ direction and periodic boundaries along the $y$ direction. The edge states (red curves) are gapped  for (a,b) $t = 0.6$. No edge states exist for (c,d) $t = 2.0$. An EP exists for $t = \lambda + \gamma = 1.9$, where a phase transition occurs. The number of unit cells along the $x$ direction is $N=20$ with $\lambda = 1.5$ and $\gamma = 0.4$.}\label{fig2}
\end{figure}
\textit{Bulk and edge states}.---The upper and lower bands $E_\pm(\mathbf{k})$ of $H(\mathbf{k})$ are two-fold degenerate \cite{SupplMaterialsNonhermitian2018}, and these bands coalesce at EPs with $ E_{\pm}(\mathbf{k}_\textrm{EP}) = 0$ for $t = \pm \lambda ~ \pm ~\gamma$ or $\pm\sqrt{\gamma^2 - \lambda^2} $. Figure 2 shows the complex energy spectra with open and periodic boundaries along the $x$ and $y$ directions, respectively. The non-Hermitian system supports gapped complex edge states for $ \abs{t} < \abs{\gamma} + \abs{\lambda} $, as shown in the red curves in Figs.~2(a) and (b). On the other hand, for $ \abs{t} > \abs{\gamma} + \abs{\lambda} $, there are no edge states [see Figs.~2(c) and (d)]. In spite of their existence, edge states can continuously be absorbed into bulk bands and therefore are not topologically protected. In fact, the bulk bands are topologically trivial, characterized by zero Chern number (see Sec.~$\textrm{\Rmnum{1}}$ in Ref.~\cite{SupplMaterialsNonhermitian2018}) over the entire range of parameters. 

\textit{Corner states}.---While the bulk bands of $H(\mathbf{k})$ are topologically trivial, the system with open-boundary conditions in the $x$ and $y$ directions hosts four zero-energy modes at its corners, as shown in Figs.~3(a-c). Moreover, these  zero-energy states are localized only at the lower-left corner [see Fig.~3(a)]. Note that the mid-gap modes can be localized at the upper-right corner if the sign of hopping amplitude $t$ is reversed (see Fig.~S1 in Ref.~\cite{SupplMaterialsNonhermitian2018}). This mid-gap-state localization at one corner results from the interplay between the symmetry $\mathcal{M}_{xy}$ and non-Hermiticity, where each corner mode is a simultaneously topological state of two intersecting non-trivial edges (see Sec.~$\textrm{\Rmnum{3}}$ in Ref.~\cite{SupplMaterialsNonhermitian2018}). Furthermore, these corner modes are topologically protected against disorder preserving $\mathcal{M}_{xy}$ symmetry and sublattice symmetry (see Sec.~$\textrm{\Rmnum{4}}$ in Ref.~\cite{SupplMaterialsNonhermitian2018}). Note that when the mirror-rotation symmetry is broken, the mid-gap modes can be localized at more than one corner, and the sites at which mode localization occurs can be diagnosed by considering the type of asymmetric hopping and non-Hermiticity in non-Hermitian SOTIs (see Sec.~$\textrm{\Rmnum{5}}$ in Ref.~\cite{SupplMaterialsNonhermitian2018}).


%
\begin{figure}[!tb]
	\centering
	\includegraphics[width=8.4cm]{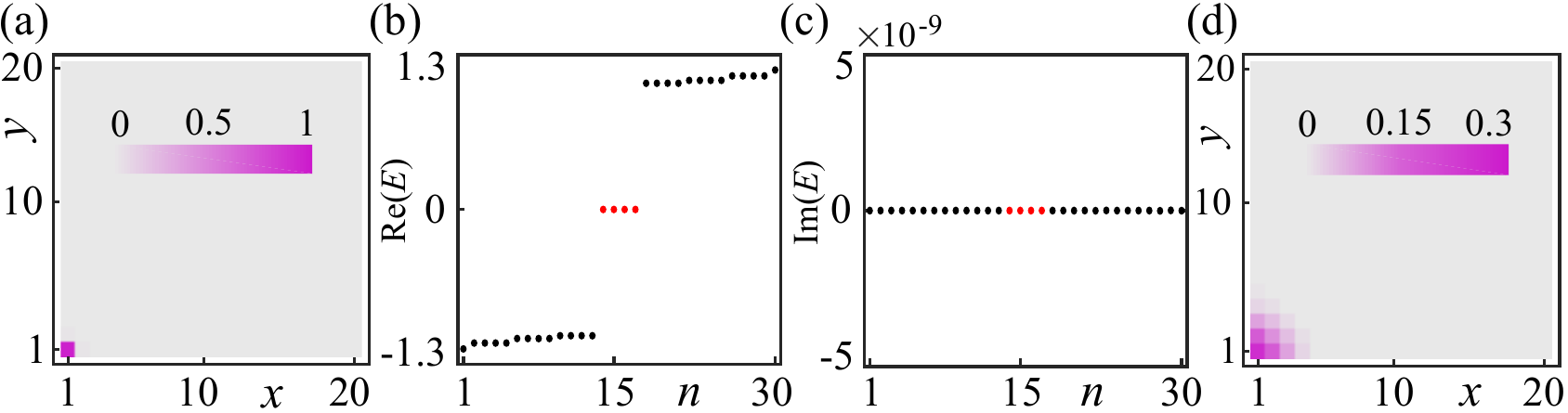}
	\caption{Corner states in the non-Hermitian SOTI described by the Hamiltonian (6). (a) Probability density distributions $\sum_{i=1}^4 \abs{\phi_{R, i, n}}^2$ ($n$ is the index of an eigenstate and $R$ specifies a unit cell) of four zero-energy states under the open boundary condition along the $x$ and $y$ directions. The zero-energy modes are localized only at the lower-left corner. (b,c) Real and imaginary parts of complex eigenenergies around zero energy. The red dots represent eigenenergies of the corner modes. The imaginary parts of the bulk eigenenergies of a finite-size sample vanish over a wide range of parameters. (d) Probability density distribution of a typical bulk state under the open boundary condition along the $x$ and $y$ directions. The bulk state is exponentially localized at the lower-left corner. The number of unit cells is $20 \times 20$ with $t = 0.6$, $\lambda = 1.5$ and $\gamma = 0.4$. }\label{fig3}
\end{figure}

Moreover, the bulk bands of the \textit{open}-boundary system are considerably different from those of the \textit{periodic} system. As shown in Figs.~3(b) and (c), the bulk eigenenergies in the case of open boundaries are entirely real over a wide range of system parameters as a consequence of pseudo-Hermiticity of the open-boundary system \cite{SupplMaterialsNonhermitian2018}, while they are complex in the case of the periodic boundaries. Furthermore, we find that, in contrast to the Hermitian SOTI, the bulk modes are exponentially localized at the lower-left corner due to the non-Hermiticity caused by the asymmetric hopping (see Sec.~$\textrm{\Rmnum{6}}$ and $\textrm{\Rmnum{7}}$ in Ref.~\cite{SupplMaterialsNonhermitian2018}), as shown in Fig.~3(d). 

\textit{Topological index}.---The topology of the non-Hermitian Hamiltonian $H_{\textrm{2D}}$ is characterized by the winding number $w$ [see Eqs.(1-5)]. One of the boundaries of the topological-phase transition calculated by this index is $t = \lambda + \gamma = 1.9$ (i.e., one of the EPs) using the parameters in Fig.~2. However, numerical calculations for the open-boundary system show that corner states exist only for $t < \sqrt{\lambda^2 + \gamma^2} \simeq 1.55$. Therefore, this topological index cannot correctly determine the phase boundary between topologically trivial and nontrivial regimes, indicating the breakdown of the usual bulk-corner correspondence in non-Hermitian systems. This breakdown results from the non-Bloch-wave behavior of open-boundary eigenstates of a non-Hermitian Hamiltonian, as studied in first-order topological insulators in Refs.~\cite{ShunyuYao2018, YaoarXiv:1804.04672}. To figure out this unexpected non-Bloch-wave behavior, \textit{complex} wavevectors, instead of real ones, are suggested for defining the topological index of non-Hermitian systems \cite{ShunyuYao2018, YaoarXiv:1804.04672}. Here we generalize this idea to the non-Hermitian SOTI (see Sec.~$\textrm{\Rmnum{7}}$ in Ref.~\cite{SupplMaterialsNonhermitian2018} for details). After replacing real wavevectors $\textbf{k}$ with complex ones 
\begin{align}\label{modifiedEq1}
\textbf{k} = (k_x, ~k_y) \to  \widetilde{\textbf{k}} = (k_x - i \textrm{ln}(\beta_0), ~ k_y - i \textrm{ln}(\beta_0)),
\end{align}
with $\beta_0 = \sqrt{\abs{(t-\gamma)/(t+\gamma)}}$, the Hamiltonian $H_{\pm}$ for $H_{\textrm{2D}}$ in Eq.~(3) has the following forms
\begin{align}\label{modifiedEq2}
\frac{\widetilde{H}_{\pm}}{\sqrt{2}}  =  \left(t - \gamma + \lambda \beta_0 \textrm{e}^{i k} \right) \sigma_{\mp} + \left(t + \gamma + \frac{\lambda}{\beta_0} \textrm{e}^{-i k} \right) \sigma_{\pm},
\end{align}
where $\sigma_{\pm} = (\sigma_x \pm i \sigma_y)/2$. Note that the location of the mid-gap corner modes depends on $\beta_0$: they are localized at the lower-left corners for $\beta_0 < 1$, and at the upper-right corners for $\beta_0 > 1$. Figure 4(a) shows the topological phase diagram. The number of zero-energy corner modes is counted as $2 \abs{w}$. Furthermore, the phase boundaries are determined by $t^2 = \lambda^2 + \gamma^2$ and $t^2 = \gamma^2 - \lambda^2$, and the phase diagram contains the trivial phase ($w=0$) and the second-order topological phase ($w=-2$).

\begin{figure}[!tb]
	\centering
	\includegraphics[width=3.6cm]{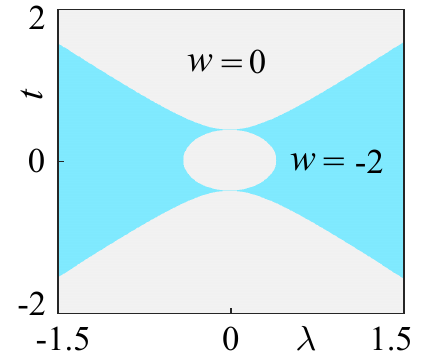}
	\caption{Topological phase diagram in the 2D non-Hermitian SOTI for $\gamma = 0.4$. The gray regions represent the topologically trivial phase with $w=0$, while the cyan regions represent the second-order topological phase with $w=-2$ that hosts corner states. The phase boundaries are determined by $t^2 = \lambda^2 + \gamma^2$ and $t^2 = \gamma^2 - \lambda^2$. }\label{fig4}
\end{figure}

\textit{3D SOTI}.---We now consider a 3D non-Hermitian Hamiltonian $H_{3D}$ that respects two-fold mirror-rotation symmetry 
\begin{equation}
{\cal M}_{xy} H_{3D} \left( k_{x}, k_{y}, k_z \right) {\cal M}_{xy}^{-1}
= H_{3D} \left( k_{y}, k_{x}, k_z \right).
\end{equation}
Note that the Hermitian counterpart was investigated in Ref.~\cite{TitusSciAdv2018}. As in the 2D case, due to the mirror-rotation symmetry in Eq.~(9), we can express the Hamiltonian $H_{3D}$ along the high-symmetry line $k_{x} = k_{y}$ as
\begin{equation}
U^{-1} H_{3D} \left( k, k, k_z \right) U = \left( \begin{array}{@{\,}cc@{\,}} 
H_{+} \left( k, k_z \right) & 0 \\
0 & H_{-} \left( k, k_z \right) \\ 
\end{array} \right),
\end{equation}
where $H_{\pm}(k, k_z)$ acts on the corresponding mirror-rotation subspace. We can define the Chern number
\begin{equation}
C_{\pm} := \frac{1}{2\pi} \int_{\textrm{BZ}} \!\! \textrm{Tr}\left[d A_{\pm} + i A_{\pm} \wedge A_{\pm}\right],
\end{equation}
where  $A_{\pm}^{\alpha \beta} = i \bra{\chi_{\pm}^{\alpha}(k, k_z)}  \ket{d \phi_{\pm}^{\beta}(k, k_z)} $ with $\alpha$ and $\beta$ taken over the filled bands, and $\ket{\phi_{\pm}^{\alpha}}$ ($\ket{\chi_{\pm}^{\alpha}}$) is a right (left) eigenstate of $H_{\pm}(k, k_z)$. This formula is a natural generalization of the single-band non-Hermitian Chern number discussed in Ref.~\cite{PhysRevLett.120.146402} to multiple bands. Then the topological index that characterizes the second-order topological phases in 3D is 
\begin{align}
C := C_{+}-C_{-}.
\end{align} 

We investigate a concrete model of a 3D non-Hermitian SOTI on a cubic lattice described by 
\begin{align}\label{3DEq1}
H_{\textrm{3D}} = & \left(m + t \sum_{i} \cos k_i \right) \tau_z + \sum_{i} \left(\Delta_1 \sin k_i + i \gamma_i\right) \sigma_i \tau_x \nonumber \\
& + \Delta_2 \left(\cos k_x - \cos k_y\right) \tau_y,  
\end{align}
where $i$ runs over $x, ~y$ and $z$, and $\gamma_x = \gamma_y = \gamma_0$. This Hamiltonian $H_{\textrm{3D}}$ only preserves mirror-rotation symmetry $\mathcal{M}_{xy}$ (see Sec.~$\textrm{\Rmnum{9}}$ in Ref.~\cite{SupplMaterialsNonhermitian2018}).

When the bulk bands of $H_{\textrm{3D}}$ are gapped and first-order-topologically trivial, it does not support gapless surface states, as shown by energy spectra with open boundaries along the $y$ direction in Figs.~5(a) and (b). However, the system with open boundaries in both $x$ and $y$ directions hosts four-fold degenerate second-order boundary modes, as shown in Figs.~5(c) and (d). In contrast to the Hermitian case \cite{TitusSciAdv2018}, these second-order boundary modes under the open boundary condition along all the directions are localized not along the hinge but anomalously localized at one corner [see Fig.~5(e)].  This indicates that the usual bulk-hinge correspondence is broken for the 3D non-Hermitian SOTI. Moreover, these second-order boundary modes are only localized at the corners on the $x=y$ plane due to the mirror-rotation symmetry  $\mathcal{M}_{xy}$ (see Fig.~S10 in Ref.~\cite{SupplMaterialsNonhermitian2018}). In addition, the second-order boundary modes can be localized at more than one corner when the mirror-rotation symmetry is broken or there exists the balanced gain and loss  (see Sec.~$\textrm{\Rmnum{9}}$ in Ref.~\cite{SupplMaterialsNonhermitian2018}).

Due to mirror-rotation symmetry, the second-order topological phase in 3D can be characterized by the Chern number $C$ [see Eqs.~(9-12)]. To generalize the bulk-boundary correspondence in 3D non-Hermitian SOTIs, we take into account the exponential-decay behavior of non-Hermitian eigenstates with open boundaries along all the directions. After considering a low-energy continuum model of the Hamiltonian $H_\textrm{3D}$ to capture the essential physics of the 3D non-Hermitian SOTI with analytical results, and replacing real wavevectors $\textbf{k}$ with complex ones (see Sec.~$\textrm{\Rmnum{9}}$ in Ref.~\cite{SupplMaterialsNonhermitian2018} for details), the Hamiltonian $H_{\pm}$ for $H_{\textrm{3D}}$ in Eq.~(10) can be expressed as
\begin{align}
\bar{H}_{\pm}(k,k_z) = & -\left[m + 3 t - t (k - i \alpha_0)^2- \frac{t}{2}\left(k_z-i \alpha_z \right)^2 \right] \sigma_z \nonumber \\
& \pm \sqrt{2}\left[\Delta_1 (k - i \alpha_0) + i  \gamma_0 \right]  \sigma_y \nonumber \\
& - \Delta_1 \left( k_z - i \alpha_z  \right) \sigma_x,
\end{align}
where 
\begin{align} 
\alpha_0 = \frac{\gamma_0}{\Delta_1},~ \textrm{and}~ \alpha_z = \frac{\gamma_z}{\Delta_1} .
\end{align}
Figure 5(f) shows the topological phase diagram, where the second-order topological phases are characterized by the nonzero Chern number ($C = -2$). The number of hinge states is counted as $2 \abs{C}$.

\begin{figure}[!tb]
	\centering
	\includegraphics[width=8.4cm]{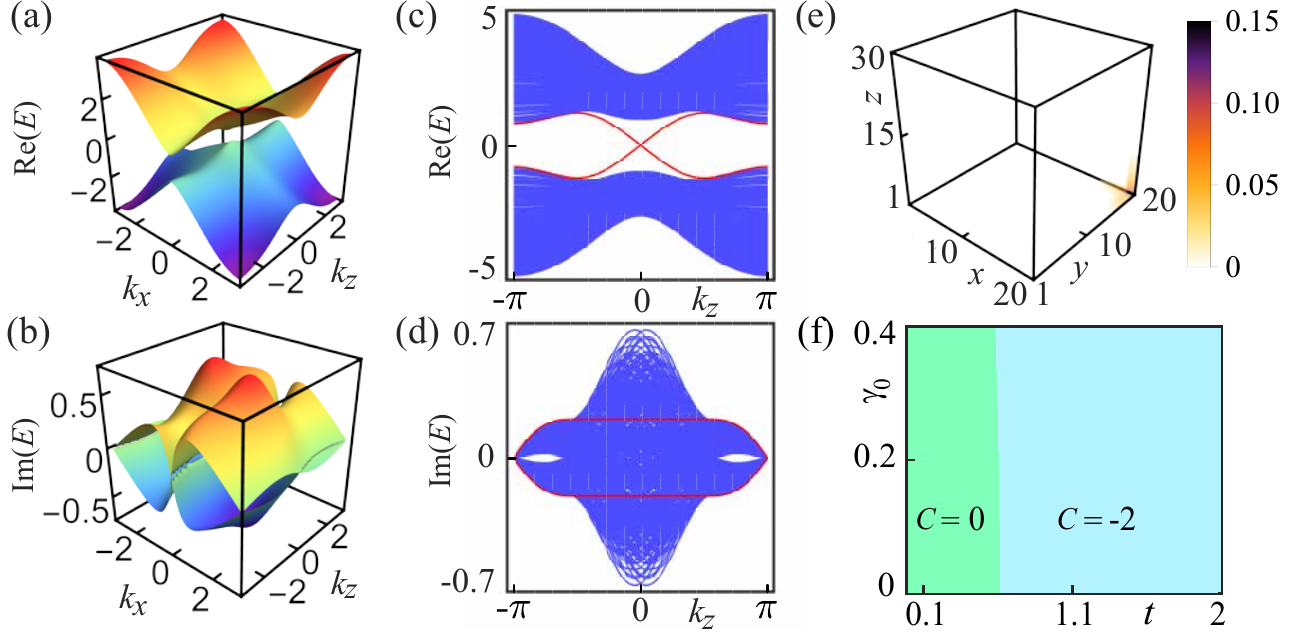}
	\caption{Three-dimensional non-Hermitian SOTI described by Eq.~(13). (a,b) Complex energy spectrum under the open boundary condition along the $y$ direction. (c,d) Complex energy spectrum under the open boundary condition along the $x$ and $y$ directions. Red curves denote four-fold degenerate second-order boundary modes. (e) Probability density distribution $\abs{\Phi_{n, R}}^2$ ($n$ is the index of an eigenstate and $R$ specifies a lattice site)  of mid-gap modes with open boundaries along the $x$, $y$ and $z$ directions. The mid-gap states (with eigenenergies of 0.035) are  localized only at one corner. The number of unit cells is 20 $\times$ 20 $\times$ 30 with $t = 1$, $\gamma_0 = 0.7$, $\gamma_z = -0.2$, $m = -2$, $\Delta_1 = 1.2$, and $\Delta_2 = 1.2$. (f) Second-order topological phase diagram characterized by the nonzero Chern number.}\label{fig5}
\end{figure}

\textit{Conclusions}.---In this Letter, we have analyzed 2D and 3D SOTIs in the presence of non-Hermiticity. In spite of their first-order topologically trivial bulk bands, second-order boundary modes exist in both 2D and 3D SOTIs. In contrast to the Hermitian cases, the mid-gap states in 2D are localized only at one corner protected by mirror-rotation symmetry and sublattice symmetry, and the second-order boundary modes are anomalously localized at a corner in 3D. The winding number (Chern number) defined by complex wavevectors is used to determine their second-order topological phases in 2D (3D). An experimental realization with ultracold atoms is also discussed. Our study provides a framework to explore richer non-Hermitian physics in higher-order topological phases.

\begin{acknowledgments}
T.L. thanks James Jun He for discussions, and Yi Peng for technical assistance. T.L. acknowledges support from a JSPS Postdoctoral Fellowship (P18023). Y.R.Z. was partially supported by China Postdoctoral Science Foundation (grant No. 2018M640055). Z.G. was supported by MEXT. K.K. acknowledges support from the JSPS through the Program for Leading Graduate Schools (ALPS). M.U. acknowledges support by  KAKENHI Grant No. JP18H01145 and a Grant-in-Aid for Scientific Research on Innovative Areas ``Topological Materials Science’’ (KAKENHI Grant No. JP15H05855) from the JSPS. F.N. is supported in part by the:
MURI Center for Dynamic Magneto-Optics via the Air Force Office of Scientific Research (AFOSR) (Grant No. FA9550-14-1-0040), Army Research Office (ARO) (Grant No. W911NF-18-1-0358), Asian Office of Aerospace Research and Development (AOARD) (Grant No. FA2386-18-1-4045), Japan Science and Technology Agency (JST) (Q-LEAP program, ImPACT program, and CREST Grant No. JPMJCR1676), JSPS (JSPS-RFBR Grant No. 17-52-50023, and JSPS-FWO Grant No. VS.059.18N), RIKEN-AIST Challenge Research Fund, and the John Templeton Foundation. 
\end{acknowledgments}

\textit{Note added}.---After this work was submitted, a related preprint \cite{arXiv:1810.11824} appeared, which focuses on the interplay between topological modes and skin boundary modes induced by non-reciprocity in non-Hermitian higher-order topological phases.


%

\clearpage \widetext
\begin{center}
	\section{Supplemental Material for ``Second-order topological phases in non-Hermitian systems"}
\end{center}
\setcounter{equation}{0} \setcounter{figure}{0}
\setcounter{table}{0} \setcounter{page}{1} \setcounter{secnumdepth}{3} \makeatletter
\renewcommand{\theequation}{S\arabic{equation}}
\renewcommand{\thefigure}{S\arabic{figure}}
\renewcommand{\bibnumfmt}[1]{[S#1]}
\renewcommand{\citenumfont}[1]{S#1}

\makeatletter
\def\@hangfrom@section#1#2#3{\@hangfrom{#1#2#3}}
\makeatother


\maketitle

\section{Tight-binding Hamiltonian}
As shown in Fig.~1 of the main text, we consider a minimal model of a second-order topological insulator on a square lattice, where each unit cell contains four sublattice degrees of freedom and asymmetric particle hoppings within each unit cell are considered. The tight-binding Hamiltonian in real-space representation is written as
\begin{align}
H_{\textrm{tb}} & = \sum_{\mathbf{R}} \left[(t - \gamma) (c_{\mathbf{R}, 1}^{\dagger} c_{\mathbf{R}, 3} + c_{\mathbf{R}, 4}^{\dagger} c_{\mathbf{R}, 2}) + (t + \gamma) (c_{\mathbf{R}, 3}^{\dagger} c_{\mathbf{R}, 1} + c_{\mathbf{R}, 2}^{\dagger} c_{\mathbf{R}, 4}) \right. \nonumber \\
&~~~ + (t - \gamma) (c_{\mathbf{R}, 3}^{\dagger} c_{\mathbf{R}, 2} - c_{\mathbf{R}, 1}^{\dagger} c_{\mathbf{R}, 4}) + (t + \gamma) (c_{\mathbf{R}, 2}^{\dagger} c_{\mathbf{R}, 3} - c_{\mathbf{R}, 4}^{\dagger} c_{\mathbf{R}, 1}) \nonumber \\
&~~~ \left. + \lambda (c_{\mathbf{R}, 1}^{\dagger} c_{\mathbf{R} + \hat{\mathbf{x}}, 3} + c_{\mathbf{R}, 4}^{\dagger} c_{\mathbf{R} + \hat{\mathbf{x}}, 2} + \textrm{H.c.})  + \lambda (c_{\mathbf{R}, 3}^{\dagger} c_{\mathbf{R} + \hat{\mathbf{y}}, 2} - c_{\mathbf{R}, 1}^{\dagger} c_{\mathbf{R} + \hat{\mathbf{y}}, 4} + \textrm{H.c.}) \right],
\end{align}
where $c_{\mathbf{R}, i}^{\dagger}$ ($i = 1, 2, 3, 4$) is the creation operator of a fermion at sublattice $i$ of unit-cell site $\mathbf{R}$, $\hat{\mathbf{x}}$ and $\hat{\mathbf{y}}$ denote the unit vectors along the $x$ and $y$ directions, $\lambda$ is the inter-cell hopping amplitude, and $t ~\pm~ \gamma$ denote the asymmetric intra-cell hopping amplitudes. In the momentum-space representation, the Hamiltonian $H_{\textrm{tb}}$ is written as $\mathcal{H} = \sum_{\mathbf{k}} \psi_{\mathbf{k}}^{\dagger} H_{\textrm{2D}}(\mathbf{k}) \psi_{\mathbf{k}}$ with $\psi_{\mathbf{k}} = (c_{\mathbf{k},1}, c_{\mathbf{k},2}, c_{\mathbf{k},3}, c_{\mathbf{k},4})^T$. Then, we have
\begin{align}
H_{\textrm{2D}}(\mathbf{k}) = \left[t + \lambda \cos(k_x)\right] \tau_x - \left[\lambda \sin(k_x) ~ + i \gamma \right] \tau_y \sigma_z + \left[t + \lambda \cos(k_y)\right] \tau_y \sigma_y + \left[\lambda \sin(k_y) + i \gamma\right] \tau_y \sigma_x, 
\end{align}
where we have set the lattice constant $a_0 = 1$, $\sigma_i$ ($i = x, y, z$) is a Pauli matrix acting on the sublattice index of particles 1 and 2 as well as particles 3 and 4 within a unit cell [see Fig.~1(a) in main text], and $\tau_i$ is a Pauli matrix acting on the space of these two pairs.

The eigenenergies of $H_{\textrm{2D}}(\mathbf{k})$ are  
\begin{align}
E_{\pm}(\mathbf{k})  = \pm \left[2 t^2 - 2 \gamma^2 + 2 \lambda^2 + 2 \lambda t \cos(k_x) + 2 \lambda t \cos(k_y) + 2 i \lambda \gamma \sin(k_x) + 2 i \lambda \gamma \sin(k_y) \right]^\frac{1}{2},
\end{align}
where each of the upper and lower energy bands is two-fold degenerate. The upper and lower bands coalesce at EPs with $ E_{\pm}(\mathbf{k}_\textrm{EP}) = 0$ for $k_x = k_y = 0 ~(\pi)$ or $k_x = -k_y$.

The bulk bands of $H_{\textrm{2D}}(\mathbf{k})$ are first-order topologically trivial in the entire range of parameters, and characterized by zero Chern number \cite{SMPhysRevLett.120.146402} defined by 
\begin{align}
\mathcal{N} = \frac{1}{2\pi} \int_{\textrm{BZ}} \!\! \textrm{Tr}[F_{xy}(\mathbf{k})] ~d^2k,
\end{align}
where the trace is taken over the occupied bands, $F_{xy}(\mathbf{k})$ is the non-Abelian Berry curvature
\begin{align}
F_{xy}^{\alpha \beta}(\mathbf{k}) = {\partial_x} A_y^{\alpha \beta}(\mathbf{k}) - {\partial_y} A_x^{\alpha \beta}(\mathbf{k}) + i [A_x, A_y]^{\alpha \beta}.
\end{align}
Here $A_{\mu}$ is the Berry connection
\begin{align}
A_{\mu}^{\alpha \beta}(\mathbf{k}) = i \bra{\chi_n^{\alpha}(\mathbf{k})}  \ket{\partial_{\mu} \phi_n^{\beta}(\mathbf{k})},
\end{align}
where $\ket{\phi_n^{\alpha}(\mathbf{k})}$ and $\ket{\chi_n^{\alpha}(\mathbf{k})}$ are the right and left eigenstates: 
\begin{align}
& H_{\textrm{2D}}(\mathbf{k}) \ket{\phi_n^{\alpha}(\mathbf{k})} = E_n \ket{\phi_n^{\alpha}(\mathbf{k})}, \\
& H_{\textrm{2D}}^{\dagger}(\mathbf{k}) \ket{\chi_n^{\alpha}(\mathbf{k})} = E_n^{*} \ket{\chi_n^{\alpha}(\mathbf{k})},
\end{align}
and $\alpha$ denotes the band degeneracy. The right and left eigenstates satisfy the following biorthogonal normalization condition
\begin{align}
\bra{\chi_n^{\alpha}(\mathbf{k})} \ket{\phi_m^{\beta}(\mathbf{k})} = \delta_{nm} \delta_{\alpha \beta}.
\end{align}
Numerical calculations show that $\mathcal{N} = 0$, indicating that the bulk bands are topologically trivial.

\section{Pseudo-Hermiticity}
In this section, we argue that the real spectrum of our non-Hermitian system, with open boundaries along both $x$ and $y$ directions, results from pseudo-Hermiticity of the real-space Hamiltonian \cite{SMPseudoHermiticity2002I, SMPseudoHermiticity2002II, SMPseudoHermiticity2002III, SMPhysRevB.84.205128}.

We rewrite the real-space Hamiltonian [see Eq.~(S1)] as $H_{\textrm{tb}} = H_{\textrm{tb}}^1 + H_{\textrm{tb}}^2 + H_{\textrm{tb}}^3 $, where
\begin{align}\label{wnEq1}
& H_{\textrm{tb}}^1 = \sum_{n_x, n_y} \Phi_{n_x, n_y}^{\dagger} \left[t (\tau_x + \tau_y \sigma_y) - i \gamma (\tau_y \sigma_z - \tau_y \sigma_x) \right] \Phi_{n_x, n_y}, \\
& H_{\textrm{tb}}^2 = \sum_{n_x, n_y} \Phi_{n_x, n_y}^{\dagger} \left[\frac{\lambda}{2}  (\tau_y \sigma_y - i \tau_y \sigma_x) \right] \Phi_{n_x, n_y + 1} + \textrm{H.c.}, \\
& H_{\textrm{tb}}^3 = \sum_{n_x, n_y} \Phi_{n_x, n_y}^{\dagger} \left[\frac{\lambda}{2}  (\tau_x + i \tau_y \sigma_z) \right] \Phi_{n_x + 1, n_y} + \textrm{H.c.}.
\end{align}
Here $n_x$ ($n_x = 1, 2, ~..., L$) and $n_y$ ($n_y = 1, 2, ..., L$) are integer-valued coordinates of unit cells in the $x$ and $y$ directions, respectively, $\sigma_i$ and $\tau_i$ ($i = x, y, z$) are Pauli matrices for the degrees of freedom within a unit cell, and $\Phi_{n_x, n_y} = (c_{n_x, n_y, A}, c_{n_x, n_y, B}, c_{n_x, n_y, C}, c_{n_x, n_y,D})^T$ is the column vector of annihilation operators with $A, B, C$, and $D$  corresponding to indexes 1, 2, 3, and 4 in Fig.~1 in the main text and denoting four orbitals within a unit cell. In the basis $\Phi = (\Phi_{1, 1}, ~\Phi_{1, 2}, ~..., ~\Phi_{L, L-1}, ~\Phi_{L, L})$, the Hamiltonian $H_\textrm{tb}$ is expressed as
\begin{align}\label{phEq1}
H_{\textrm{tb}} =  \Phi^{\dagger} H_0 \Phi,
\end{align}
where $H_0$ is the matrix form of the Hamiltonian $H_{\textrm{tb}}$.

\begin{figure}[!tb]
	\centering
	\includegraphics[width=8cm]{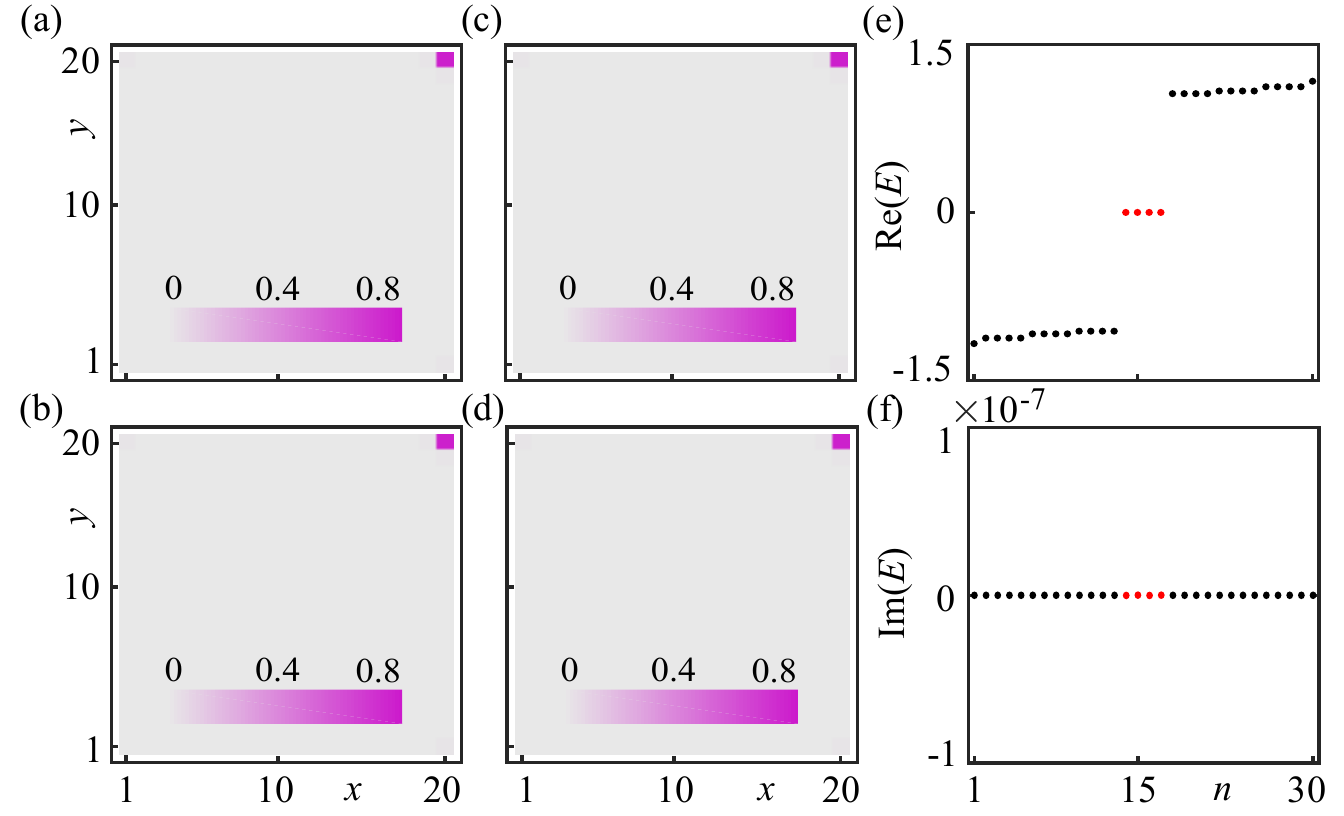}
	\caption{(a-d) Probability density distributions $\sum_{i}^{4} \abs{\phi_{R, i, n}}^2$ ($n$ is the index of an eigenstate and $R$ specifies a unit cell) of four zero-energy states under the open boundary condition along both $x$ and $y$ directions for the $20 \times 20$ unit cells with $t=-0.6$, $\lambda=1.5$, and $\gamma = 0.4$. All the zero-energy states are localized at the upper-right corner. (e, f) Real and imaginary parts of complex eigenenergies close to zero energy. The red dots represent the eigenenergies of the corner modes. The bulk eigenenergies for a finite-size sample are real over a wide range of parameters. }\label{figSM1}
\end{figure}

\begin{figure}[!tb]
	\centering
	\includegraphics[width=8cm]{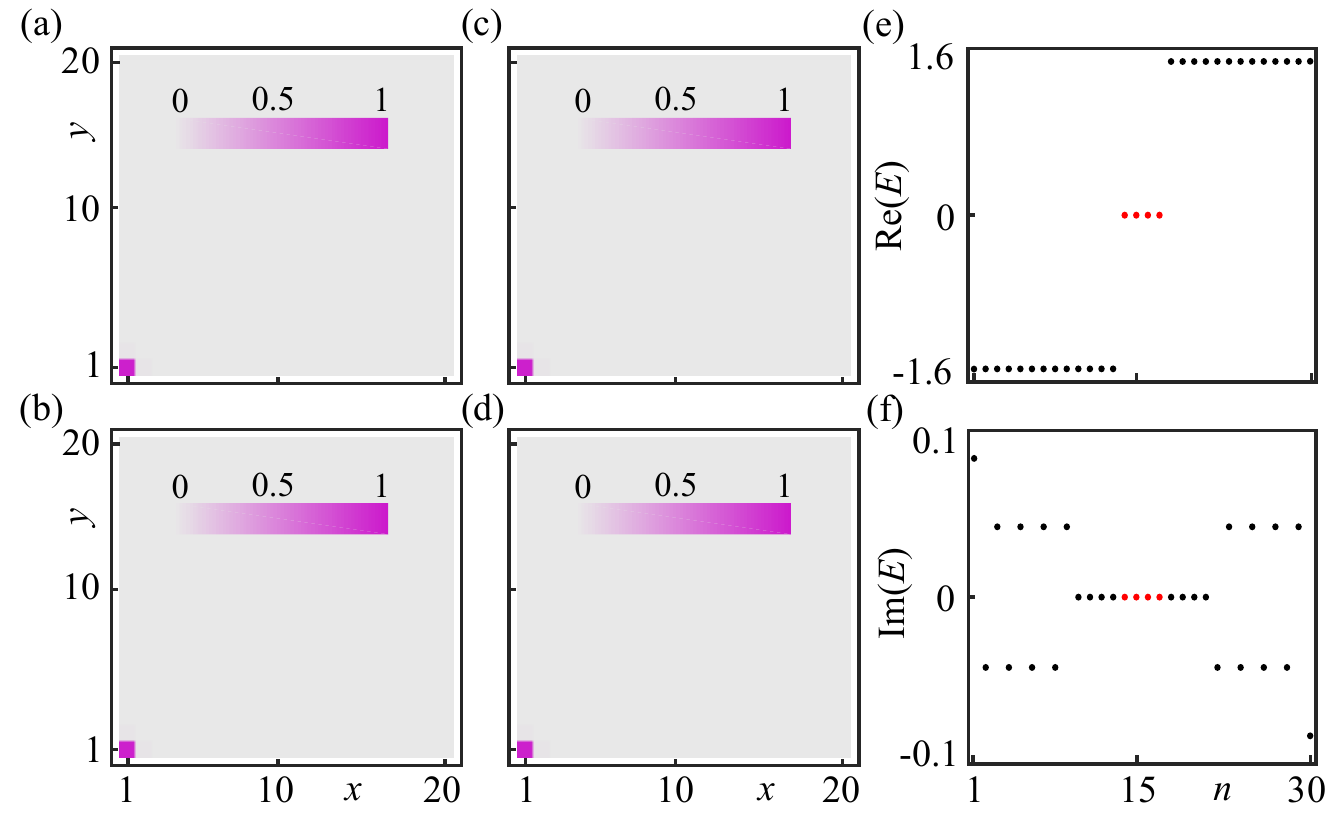}
	\caption{(a-d) Probability density distributions of four zero-energy states under the open boundary condition along both $x$ and $y$ directions for the $20 \times 20$ unit cells with $t=0.3$, $\lambda=1.5$, and $\gamma = 0.4$. All the zero-energy states are localized at the lower-left corner. ~ (e, f) Real and imaginary parts of complex eigenenergies close to zero energy. The red dots represent the eigenenergies of the corner modes. The bulk eigenenergies for a finite-size sample can be complex for the parameters considered here.}\label{figSM2}
\end{figure}

The Hamiltonian $H_\textrm{tb}$ is pseudo-Hermitian, which satisfies
\begin{align}\label{phEq2}
\eta H_0^{\dagger} \eta^{-1} = H_0,
\end{align}
where $\eta$ is a $4L \times 4L$ square matrix, and only contains elements $\sigma_y$ at its anti-diagonal sites: 
\begin{align}
\eta = \left[\begin{matrix}
0 & 0 & \dots & 0 & \sigma_y \\
0 & 0 & \dots & \sigma_y & 0 \\
\vdots & \vdots & \reflectbox{$\ddots$} & \vdots & \vdots \\
0 & \sigma_y & \dots & 0 & 0 \\
\sigma_y & 0 & \dots & 0 & 0  \\  \end{matrix}\right].
\end{align}
While positivity is usually required for the definition of pseudo-Hermiticity \cite{SMPseudoHermiticity2002I, SMPseudoHermiticity2002II, SMPseudoHermiticity2002III}, we do not assume the positivity here. From Eq.~(\ref{phEq2}), for any eigenenergy $E_n$ with the eigenequation $H_0 \ket{\phi_n} = E_n \ket{\phi_n}$, we have 
\begin{align}\label{phEq3}
E_n \bra{\phi_n} \eta^{-1} \ket{\phi_n} = E_n^{*} \bra{\phi_n} \eta^{-1} \ket{\phi_n}.
\end{align}
Therefore, for $\bra{\phi_n} \eta^{-1} \ket{\phi_n} \neq 0$, we have real eigenenergy $E_n$ for the open-boundary systems, which holds for a wide range of parameters (see Fig.~\ref{figSM1} and Fig.~3 in the main text). Note that the bulk eigenenergies can be complex in a certain range of parameters, as shown in Fig.~\ref{figSM2}.

\begin{figure}[!tb]
	\centering
	\includegraphics[width=4cm]{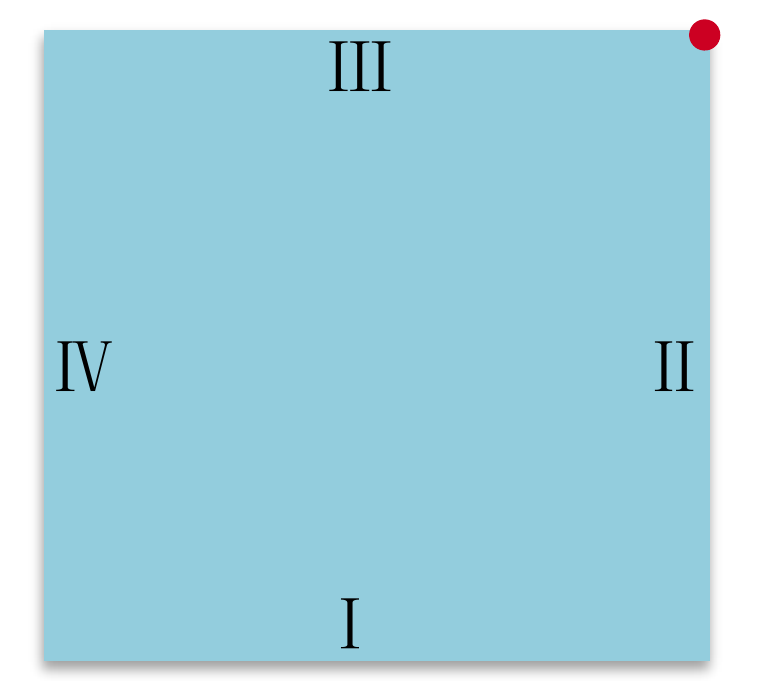}
	\caption{Schematic illustration showing a 2D non-Hermitian SOTI in a square sample. All four zero-energy modes are localized at the upper-right corner for $t < 0 $, and $\gamma > 0$. $\textrm{\Rmnum{1}}$, $\textrm{\Rmnum{2}}$, $\textrm{\Rmnum{3}}$ and $\textrm{\Rmnum{4}}$ label the four edges of the lattice.}\label{figSM3}
\end{figure}

\section{Edge theory}
As we argue in the main text, the mid-gap-state localization at one corner results from the interplay between symmetry $\mathcal{M}_{xy}$ and non-Hermiticity, where each corner mode is a mutual topological state of two intersecting nontrivial edges. In this section, we develop an edge theory to explain this result. We label the four edges of a square sample as $\textrm{\Rmnum{1}}, \textrm{\Rmnum{2}}, \textrm{\Rmnum{3}}~ \textrm{and}~ \textrm{\Rmnum{4}}$ (see Fig.~\ref{figSM3}). For the sake of simplicity, we consider the case of $\min\{\lambda, \, -t, \, \lambda + t, \, \gamma\} > 0$, $\lambda \gg \max\{\gamma, \, \lambda + t\}$, and $\gamma > \lambda + t $. In this case, the low-energy edge bands of the Hermitian part of $H_{\textrm{2D}}(\mathbf{k})$ lie around the $\Gamma$ point of the Brillouin zone. Therefore, we consider the continuum model of the lattice Hamiltonian [see Eq.~(S2)] by expanding its wavevector $\mathbf{k}$ to first-order around the $\Gamma = (0, 0)$ point of the Brillouin zone, obtaining
\begin{align}\label{edgeEq1}
\tilde{H}_{\textrm{cm}} =  \left(t + \lambda \right) \tau_x - \left(\lambda k_x + i \gamma \right)  \tau_y \sigma_z + \left(t + \lambda  \right) \tau_y \sigma_y + \left(\lambda k_y  + i \gamma \right) \tau_y \sigma_x.
\end{align}

We first investigate the edge $\textrm{\Rmnum{1}}$ of the four edges. Substituting $k_y$ by $-i \partial_{y}$, and treating terms including $ t + \lambda $ and $\gamma$ as perturbations (which are valid if they are relatively small), we can rewrite the Hamiltonian $ \bar{H}_{\textrm{cm}} $ into the sum of the following two terms:
\begin{align}\label{edgeEq2}
\tilde{H}_{\textrm{cm},1} & =  \left(t + \lambda  \right) \tau_y \sigma_y -i \lambda \frac{\partial}{\partial y}   \tau_y \sigma_x, \\
\tilde{H}_{\textrm{cm},2} & =  \left(t + \lambda \right) \tau_x - \left(\lambda k_x + i \gamma \right)  \tau_y \sigma_z + i \gamma  \tau_y \sigma_x,
\end{align}
where $\tilde{H}_{\textrm{cm},1}$ is Hermitian, and $\tilde{H}_{\textrm{cm},2}$ is treated as the perturbation for $\lambda \gg \gamma,\, t + \lambda$.
To solve the eigenvalue equation $\tilde{H}_{\textrm{cm},1} \phi_{\textrm{cm}}^{\textrm{\Rmnum{1}}} (y) = E_{\textrm{cm}} \phi_{\textrm{cm}}^{\textrm{\Rmnum{1}}} (y)$ with $E_{\textrm{cm}} = 0$ under the boundary condition $\phi_{\textrm{cm}}^{\textrm{\Rmnum{1}}}(+\infty) =0$, we can write the solution in the following form
\begin{align}\label{edgeEq3}
\phi_{\textrm{cm}}^{\textrm{\Rmnum{1}}} (y) = \mathcal{N}_y \exp(-\alpha_{\textrm{\Rmnum{1}}} y) \exp(i k_x x) \chi_{\textrm{\Rmnum{1}}}, ~~~ \textrm{Re}(\alpha_{\textrm{\Rmnum{1}}}) > 0,
\end{align}
where $\mathcal{N}_y$ is a normalization constant. The eigenvector $\chi_{\textrm{\Rmnum{1}}}$ satisfies $\sigma_z \chi_{\textrm{\Rmnum{1}}} =  - \chi_{\textrm{\Rmnum{1}}}$ with
\begin{align}\label{edgeEq4}
\ket{\chi_{\textrm{\Rmnum{1}}}^1} &= \ket{\tau_z = 1} \otimes \ket{\sigma_z = -1}, \\
\ket{\chi_{\textrm{\Rmnum{1}}}^2} &= \ket{\tau_z = -1} \otimes \ket{\sigma_z = -1} .
\end{align}
Then, the effective Hamiltonians for the edge $\textrm{\Rmnum{1}}$ can be obtained in this basis as
\begin{align}\label{edgeEq5}
\mathcal{H}_{\textrm{edge}}^{\textrm{\Rmnum{1}}}  = \int_{0}^{+\infty} \!\! \left(\phi_{\textrm{cm}}^{\textrm{\Rmnum{1}}}\right) ^{*}\!(y) \; \tilde{H}_{\textrm{cm},2} \; \phi_{\textrm{cm}}^{\textrm{\Rmnum{1}}} (y) \: dy .
\end{align}
Therefore, we have
\begin{align}\label{edgeEq6}
\mathcal{H}_{\textrm{edge}}^{\textrm{\Rmnum{1}}} = \left(t + \lambda \right) \varrho_x + \left(\lambda k_x + i\gamma \right) \varrho_y,
\end{align}
where $\varrho_i$ ($i = x, y, z$) are Pauli matrices.

The effective Hamiltonian for the edges $\textrm{\Rmnum{2}}$, $\textrm{\Rmnum{3}}$ and $\textrm{\Rmnum{4}}$ can be obtained through similar procedures:
\begin{align}\label{edgeEq7}
\mathcal{H}_{\textrm{edge}}^{\textrm{\Rmnum{2}}} & = - \left(t + \lambda \right) \varrho_x + \left(\lambda k_y + i\gamma \right) \varrho_y, \\
\mathcal{H}_{\textrm{edge}}^{\textrm{\Rmnum{3}}} & = \left(t + \lambda \right) \varrho_x - \left(\lambda k_x + i\gamma \right) \varrho_y, \\
\mathcal{H}_{\textrm{edge}}^{\textrm{\Rmnum{4}}} & = \left(t + \lambda \right) \varrho_x + \left(\lambda k_y + i\gamma \right) \varrho_y.
\end{align}

It is straightforward to verify that the two zero-energy bound states for edges $\textrm{\Rmnum{1}}$ and $\textrm{\Rmnum{3}}$ are localized at their right ends for $\gamma > t + \lambda$ (note that each edge exhibits a zero-energy bound state for small $\gamma$), while the two zero-energy bound states for edges $\textrm{\Rmnum{2}}$ and $\textrm{\Rmnum{4}}$ are localized at their upper ends. Therefore, the zero-energy states are localized at the upper-right corner for $t < 0 $ and $\gamma > 0$ (see Fig.~S1).

\section{Robustness against disorder}
\begin{figure}[!tb]
	\centering
	\includegraphics[width=11cm]{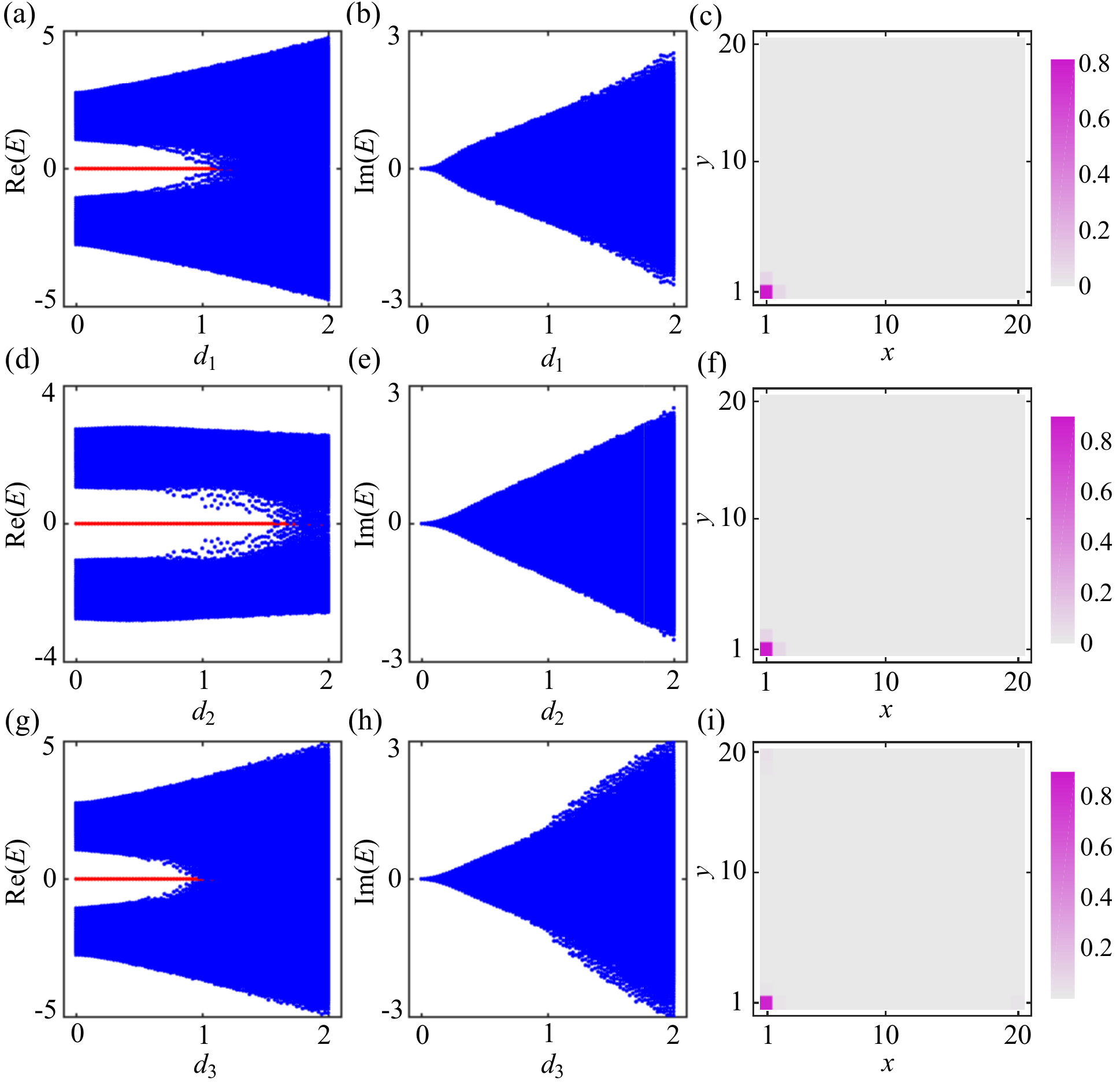}
	\caption{Energy spectra and probability density distributions under the open boundary condition in both $x$ and $y$ directions (a-c) as a function of the disorder strength $d_1$ with $d_2 = 0$ and $d_3 = 0$, (d-f) as a function of the disorder strength $d_2$ with $d_1 = 0$ and $d_3 = 0$, and (g-i) as a function of the disorder strength $d_3$ with $d_1 = 0$ and $d_2 = 0$. The other parameters are chosen to be $t = 0.6$, $\lambda = 1.5$, and $\gamma = 0.4$. (a, d, g) Real and (b, e, h) imaginary parts of the spectra. Red dots denote zero-energy modes. (c, f, i) Averaged probability density distributions of the four zero-energy states with $d_1 = 0.8$, $d_2 = 0.8$, and $d_3 = 0.8$,  respectively. }\label{figSM4}
\end{figure}

\begin{figure}[!tb]
	\centering
	\includegraphics[width=11.5cm]{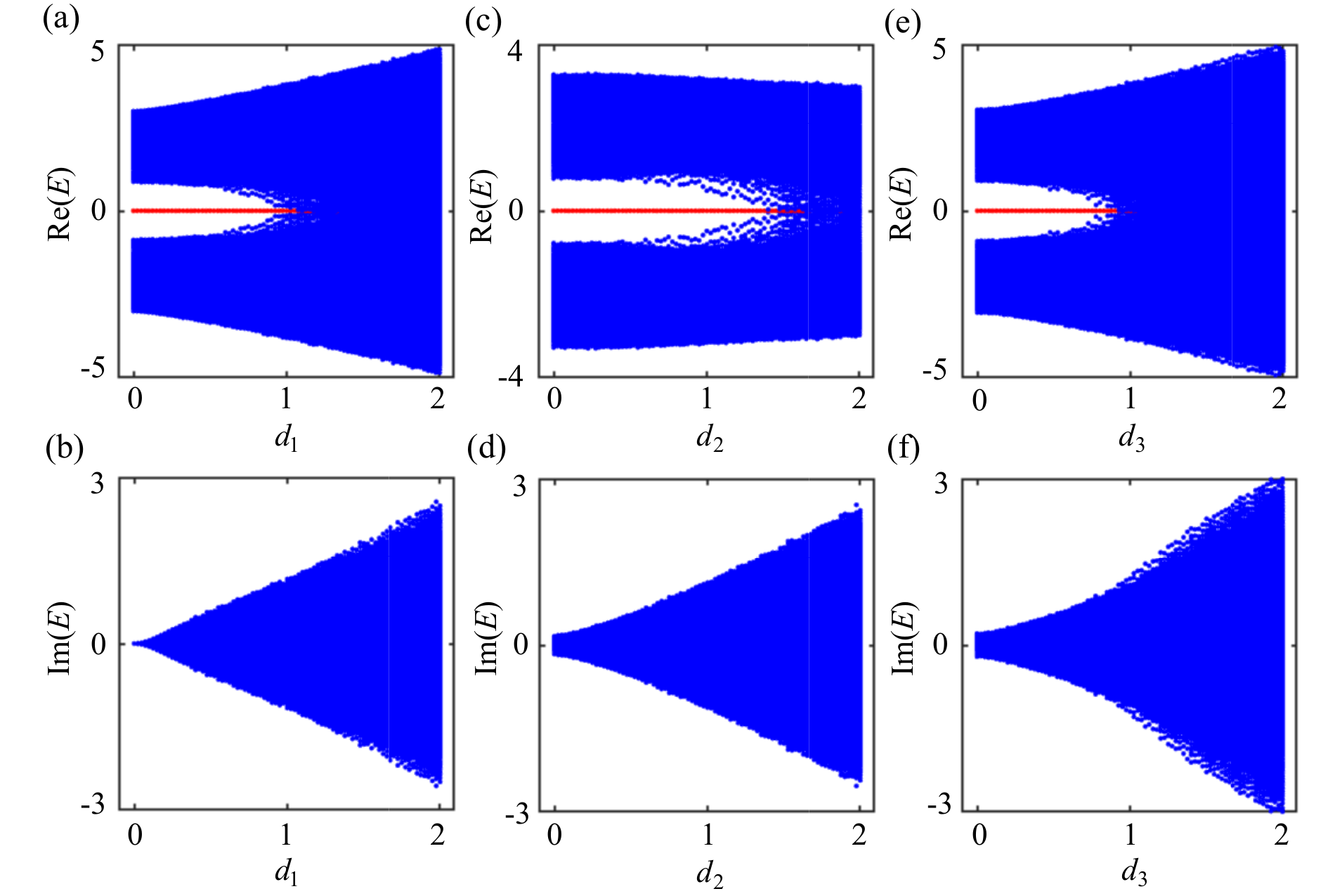}
	\caption{Energy spectra similar to those of Fig.~\ref{figSM4} but with the different values of disorder strength. (a, b) $d_2 = 0.4$ and $d_3 = 0.4$, (c, d) $d_1 = 0.4$ and $d_3 = 0.4$, and (e, f) $d_1 = 0.4$ and $d_2 = 0.4$. The corner modes are topologically protected against disorder that preserves $\mathcal{M}_{xy}$ symmetry and sublattice symmetry.}\label{figSM5}
\end{figure}

We now show that the zero-energy corner states are robust against disorder that preserves  $\mathcal{M}_{xy}$ symmetry and sublattice symmetry. We consider the following real-space disordered Hamiltonian:
\begin{align}\label{DisorderEq1}
& \bar{H}_{\textrm{tb}}^1 = \sum_{n_x, n_y} \Phi_{n_x, n_y}^{\dagger} \left[\left(t + d_1 \xi_{n_x, n_y} \right) (\tau_x + \tau_y \sigma_y) - i \left(\gamma + d_2 \zeta_{n_x, n_y}\right) (\tau_y \sigma_z - \tau_y \sigma_x) \right] \Phi_{n_x, n_y}, \\
& \bar{H}_{\textrm{tb}}^2 = \sum_{n_x, n_y} \Phi_{n_x, n_y}^{\dagger} \left[\frac{1}{2} \left(\lambda + d_3 \mu_{n_x, n_y}\right)  (\tau_y \sigma_y - i \tau_y \sigma_x) \right] \Phi_{n_x, n_y + 1} + \textrm{H.c.}, \\
& \bar{H}_{\textrm{tb}}^3 = \sum_{n_x, n_y} \Phi_{n_x, n_y}^{\dagger} \left[\frac{1}{2} \left(\lambda + d_3 \mu_{n_x, n_y}\right) (\tau_x + i \tau_y \sigma_z) \right] \Phi_{n_x + 1, n_y} + \textrm{H.c.},
\end{align}
where $\xi_{n_x, n_y}$, $\zeta_{n_x, n_y}$, and $\mu_{n_x, n_y}$ are uniform random variables distributed over $[-1, 1]$, while $d_1$, $d_2$, and $d_3$ are the corresponding disorder strength. As shown in Fig.~\ref{figSM4} and Fig.~\ref{figSM5} for the different values of disorder strength, the corner modes are topologically protected against disorder with the $\bar{\cal M}_{xy}$ symmetry and sublattice symmetry, unless the band gaps close. Moreover, the corner modes are well localized at one corner of a square sample  [see Fig.~\ref{figSM4}(c, f, i)]. 


\section{Effect of a different type of asymmetric hopping and non-Hermiticity on the localization of corner modes}

\begin{figure}[!tb]
	\centering
	\includegraphics[width=16cm]{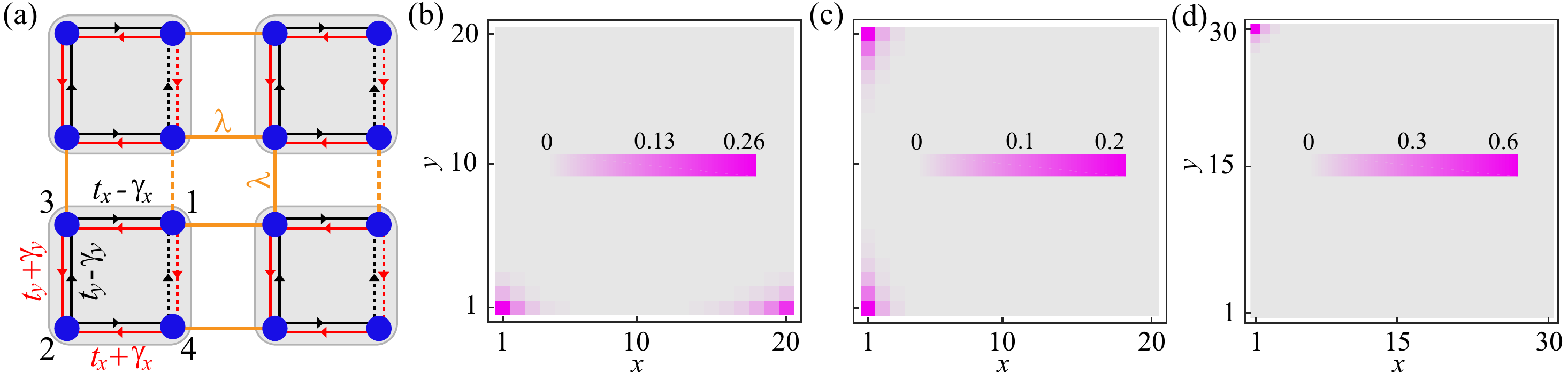}
	\caption{(a) Real-space representation of the 2D model in Eq.~(\ref{BMRS1}) on a square lattice. In contrast to the model shown in Fig.~1(a) in the main text, here we consider different amplitudes of asymmetric hopping along the $x$ and $y$ directions i.e., $\gamma_x \neq \gamma_y $. Probability density distributions of mid-gap states under the open boundary condition along the $x$ and $y$ directions: (b) for  $\gamma_x = 0$ and $\gamma_y = 0.3$, (c) for $\gamma_x = 0.3$ and $\gamma_y = 0$, and (d) for $\gamma_x = 0.3$ and $\gamma_y = -0.3$. The number of unit cells is 20 $\times$ 20 with $t_x = t_y = 1.0$ and $\lambda = 1.5$.}\label{figSM15}
\end{figure}
\begin{figure}[!tb]
	\centering
	\includegraphics[width=9cm]{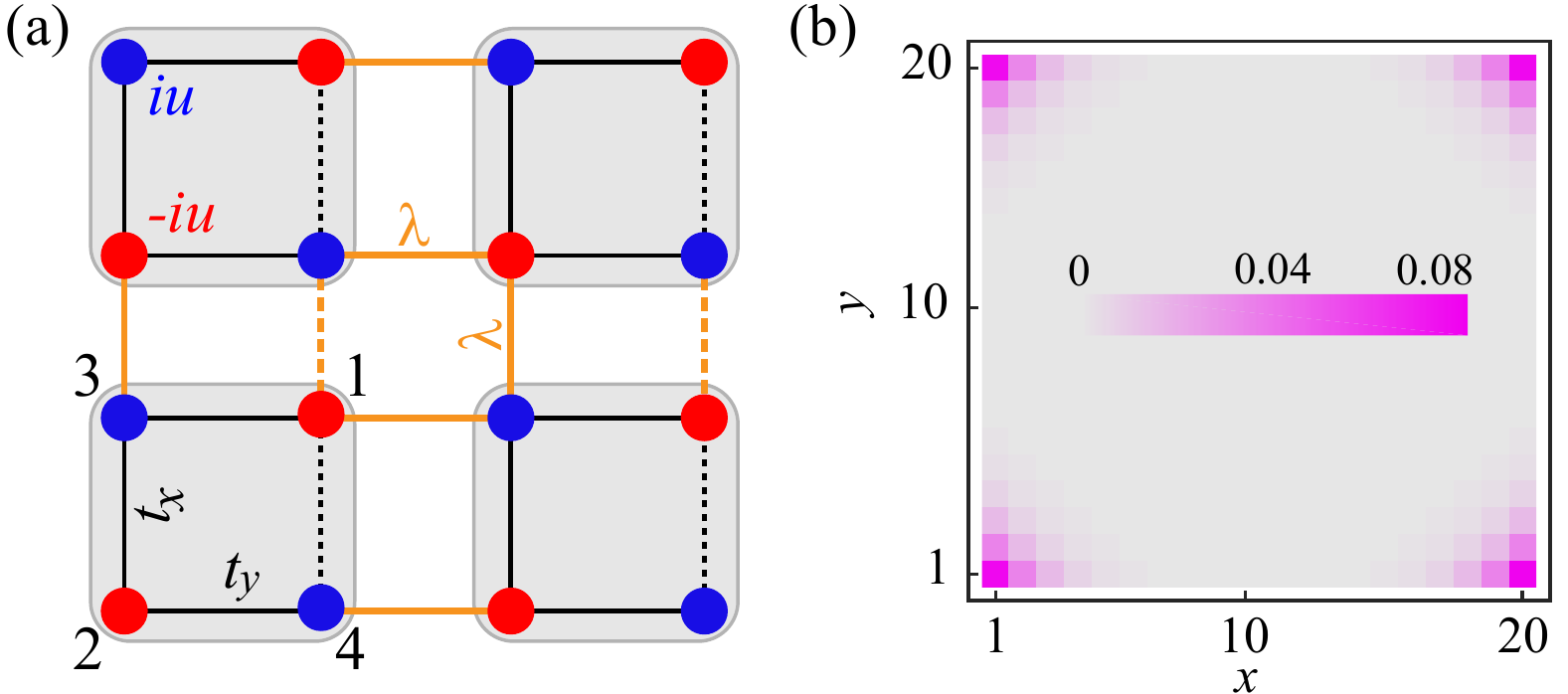}
	\caption{(a) Real-space representation of the 2D model in Eq.~(\ref{BMRS2}) in the presence of alternating on-site gain and loss (with the imaginary staggered potentials $iu$ and $-iu$) within each unit cell.  The dashed lines indicate hopping terms with a negative sign.  Probability density distributions of mid-gap states under the open boundary condition along the $x$ and $y$ directions. The number of unit cells is 20 $\times$ 20 with $u = 0.4$, $t_x = t_y = 1.0$ and $\lambda = 1.5$.}\label{figSM16}
\end{figure}

The interplay between the mirror-rotation symmetry and non-Hermiticity leads to the localization of the mid-gap states only at one corner, as explained in the main text. However, the mid-gap modes can be localized at more than one corner when the mirror-rotation symmetry is broken. In this section, we consider the case where the mirror-rotation symmetry is broken, and study the effect of a different type of asymmetric
hopping and non-Hermiticity (i.e., balanced gain and loss) on the mode localization of the mid-gap states in non-Hermitian SOTIs. Note that the effect of asymmetric hopping on the higher-order boundary modes in non-reciprocal systems has been investigated in Ref.~\cite{SMarXiv:1810.11824}.

We investigate the same lattice mode of the 2D SOTI as the one in the main text. But here we consider different amplitudes of asymmetric hopping along the $x$ and $y$ directions within each unit cell, i.e., $\gamma_x \neq \gamma_y $, as shown in Fig.~\ref{figSM15}(a). The Bloch Hamiltonian is written as
\begin{align}\label{BMRS1}
\bar{H}_{\textrm{2D}}(\mathbf{k}) = \left[t_x + \lambda \cos(k_x)\right] \tau_x - \left[\lambda \sin(k_x) ~ + i \gamma_x \right] \tau_y \sigma_z + \left[t_y + \lambda \cos(k_y)\right] \tau_y \sigma_y + \left[\lambda \sin(k_y) + i \gamma_y \right] \tau_y \sigma_x, 
\end{align}
where $\bar{H}_{\textrm{2D}}(\mathbf{k})$ only respects sublattice symmetry with $\tau_z \bar{H}_{\textrm{2D}}(\mathbf{k}) \tau_z^{-1} = - \bar{H}_{\textrm{2D}}(\mathbf{k})$ for $\gamma_x \neq \gamma_y $ for $\gamma_x \neq \gamma_y $. Note that 2D Hermitian SOTIs can exist even in the absence of crystalline symmetries \cite{SMPhysRevLett.119.246401}.

Figure \ref{figSM15}(b) shows the probability density distribution of mid-gap states for the larger hopping amplitude with $(t_y + \gamma_y)$ along the $y$ direction and the symmetric hopping along the $x$ direction [see Fig.~\ref{figSM15}(a)]. In this case, the mid-gap states are localized at both the lower-left and lower-right corners. In contrast, for the larger hopping amplitude with $(t_x + \gamma_x)$ along the $x$ direction and the symmetric hopping along the $y$ direction, the mid-gap states are localized at both the lower-left and upper-left corners, as shown in Fig.~\ref{figSM15}(c). Moreover, the mid-gap modes are localized only at the upper-left corner for the larger asymmetric hopping amplitudes with $(t_x + \gamma_x)$ and $(t_y - \gamma_y)$ along the $x$ and $y$ directions, respectively, as shown in Fig.~\ref{figSM15}(d). Therefore, the localization of corner states relies on the type of asymmetric hopping for 2D non-reciprocal lattice models. 

In addition to the different type of asymmetric hopping, the second-order boundary modes in 2D systems can be localized at more than one corner in the presence of a different type of non-Hermiticity i.e., balanced gain and loss. As shown in Fig.~\ref{figSM16}(a), we consider the alternating on-site gain and loss with symmetric particle hopping, where the imaginary staggered potentials are indicated by $iu$ and $-iu$. The Bloch Hamiltonian is written as
\begin{align}\label{BMRS2}
\tilde{H}_{\textrm{2D}}(\mathbf{k}) =  \left[t_x + \lambda \cos(k_x)\right] \tau_x - \lambda \sin(k_x)  \tau_y \sigma_z  + \left[t_y + \lambda \cos(k_y)\right] \tau_y \sigma_y + \lambda \sin(k_y)  \tau_y \sigma_x - iu \tau_z. 
\end{align}
Whereas $\tilde{H}_{\rm 2D} \left( {\bf k} \right)$ does not respect sublattice symmetry, it still respects pseudo-anti-Hermiticity  with $\tau_z \tilde{H}_{\textrm{2D}}^\dagger(\mathbf{k}) \tau_z^{-1} = - \tilde{H}_{\textrm{2D}}(\mathbf{k})$ \cite{SMPhysRevB.84.205128}. As a result, the second-order topological phase survives even in the presence of the balanced gain and loss. Figure \ref{figSM16}(b) shows the probability density distribution. In the presence of the balanced gain and loss, the mid-gap states are localized at four corners in 2D systems. In contrast to non-Hermitian SOTIs with asymmetric hopping, the eigenenergies of mid-gap states have nonzero imaginary parts as in the case for the non-Hermitian Su-Schrieffer-Heeger model with the balanced gain and loss \cite{SMPhysRevB.84.205128}.

\section{Degenerate perturbation theory}
The bulk-state localization results from the non-Hermiticity due to asymmetric hopping, which can intuitively be explained using degenerate perturbation theory when $\gamma$ is small. In this section, we consider a continuum model of the lattice Hamiltonian [see Eq.~(S2)] by expanding its wavevector $\mathbf{k}$ up to second order around the $\Gamma = (0, 0)$ point of the Brillouin zone, obtaining $H_{\textrm{cm}} = H_{\textrm{cm}}^1 + H_{\textrm{cm}}^2 $, where
\begin{align}\label{Cmodel}
H_{\textrm{cm}}^1 = & \left[t + \lambda - \frac{\lambda}{2} k_x^2 \right] \tau_x - \lambda k_x  \tau_y \sigma_z + \left[t + \lambda - \frac{\lambda}{2} k_y^2 \right] \tau_y \sigma_y + \lambda k_y  \tau_y \sigma_x, \\
H_{\textrm{cm}}^2 = & - i \gamma  \tau_y \sigma_z +  i \gamma \tau_y \sigma_x.
\end{align}
Note that $ H_{\textrm{cm}}^2 $ can be obtained from $ H_{\textrm{cm}}^1 $ as $H_{\textrm{cm}}^2 = \bar{H}_{\textrm{cm}}^2 + \bar{\bar{H}}_{\textrm{cm}}^2$:
\begin{align}\label{Cmodel2}
\bar{H}_{\textrm{cm}}^2 & =  \frac{i \gamma}{\lambda} \left(\frac{\partial H_{\textrm{cm}}^1}{\partial k_x} + \frac{\partial H_{\textrm{cm}}^1}{\partial k_y}\right) = \frac{\gamma}{\lambda}  \left([x, H_{\textrm{cm}}^1] + [y, H_{\textrm{cm}}^1]\right), \\
\bar{\bar{H}}_{\textrm{cm}}^2 & =  i \gamma \left(k_x \tau_x +  k_y \tau_y \sigma_y\right).
\end{align}

The degenerate bulk states for the Hermitian part $ H_{\textrm{cm}}^1 $ of the  Hamiltonian are denoted by $\ket{\phi_i^0}$ ($ i = 1, 2, 3, 4 $), and the non-Hermitian part  $ H_{\textrm{cm}}^2 $ is considered as a perturbation for $ \lambda \gg \gamma$, and $\mathbf{k}$ is around the $\Gamma$ point. By applying degenerate perturbation theory \cite{SMSakurai2011, SMCappellaro2012}, the first-order correction to the wavefunctions $\ket{\phi_i^0}$ for small $\mathbf{k}$ is
\begin{align}\label{Pertubation1}
\ket{\phi_i^1} = &\sum_{h \notin \{\phi_i^0\}} \left(\frac{\bra{h} H_{\textrm{cm}}^2 \ket{\phi_i^0} }{E_d^0 - E_h^0} \ket{h}  + \sum_{j \neq i}   \frac{\bra{\phi_j^0} H_{\textrm{cm}}^2 \ket{h} }{u_i - u_j} \ket{\phi_j^0} \frac{\bra{h} H_{\textrm{cm}}^2 \ket{\phi_i^0} }{E_d^0 - E_h^0} \right) \nonumber \\
\simeq &\sum_{h \notin \{\phi_i^0\}} \left(\frac{\bra{h} \bar{H}_{\textrm{cm}}^2 \ket{\phi_i^0} }{E_d^0 - E_h^0} \ket{h}  + \sum_{j \neq i}   \frac{\bra{\phi_j^0} \bar{H}_{\textrm{cm}}^2 \ket{h} }{u_i - u_j} \ket{\phi_j^0} \frac{\bra{h} \bar{H}_{\textrm{cm}}^2 \ket{\phi_i^0} }{E_d^0 - E_h^0} \right) \nonumber \\
= &\sum_{h \notin \{\phi_i^0\}} \left(\frac{\gamma}{\lambda}\ket{h} \bra{h} x \ket{\phi_i^0}    + \sum_{j \neq i}   \frac{\gamma^2 (E_d^0 - E_h^0)  \ket{\phi_j^0} \bra{h} x \ket{\phi_i^0} \bra{\phi_j^0} x \ket{h} }{\lambda^2 (u_i - u_j)}   ~+ \right. \nonumber \\
& ~~~~~~~~~~~~~~~ \left. \frac{\gamma}{\lambda}\ket{h} \bra{h} y \ket{\phi_i^0}    + \sum_{j \neq i}   \frac{\gamma^2 (E_d^0 - E_h^0)  \ket{\phi_j^0} \bra{h} y \ket{\phi_i^0} \bra{\phi_j^0} y \ket{h} }{\lambda^2 (u_i - u_j)}   \right) \nonumber \\
= & ~ \frac{\gamma}{\lambda} \left(x- \bar{x}_i \right) \ket{\phi_i^0}    + \sum_{j \neq i}   \frac{\gamma^2 (E_d^0 - E_h^0)   \left[\bra{\phi_j^0} x^2 \ket{\phi_i^0} - \bar{x}_j^{*} \bar{x}_i  \right] \ket{\phi_j^0} }{\lambda^2 (u_i - u_j)}   ~+  \nonumber \\
& ~  \frac{\gamma}{\lambda} \left(y - \bar{y}_i\right) \ket{\phi_i^0}    + \sum_{j \neq i}   \frac{\gamma^2 (E_d^0 - E_h^0)   \left[\bra{\phi_j^0} y^2 \ket{\phi_i^0} - \bar{y}_j^{*} \bar{y}_i  \right] \ket{\phi_j^0} }{\lambda^2 (u_i - u_j)}   \nonumber \\
\simeq & ~ \left[\frac{\gamma}{\lambda} \left(x- \bar{x}_i \right)    + \sum_{j \neq i}   \frac{\gamma^2 (E_d^0 - E_h^0)   \left[\bra{\phi_i^0} x^2 \ket{\phi_j^0} - \bar{x}_i^{*} \bar{x}_j  \right]  }{\lambda^2 (u_j - u_i)} \right] \ket{\phi_i^0}  ~+  \nonumber \\
& ~  \left[\frac{\gamma}{\lambda} \left(y- \bar{y}_i \right)    + \sum_{j \neq i}   \frac{\gamma^2 (E_d^0 - E_h^0)   \left[\bra{\phi_i^0} y^2 \ket{\phi_j^0} - \bar{y}_i^{*} \bar{y}_j  \right]  }{\lambda^2 (u_j - u_i)} \right] \ket{\phi_i^0} ,
\end{align}
where  $E_d^0$, $E_h^0$, $u_i$, $\bar{x}_i$ and $\bar{y}_i$ satisfy 
\begin{align}\label{Pertubation2}
& H_{\textrm{cm}}^1 \ket{\phi_i^0} =  E_d^0 \ket{\phi_i^0},  \\
& H_{\textrm{cm}}^1 \ket{h} =  E_h^0 \ket{h}, ~~~ h \notin \{\phi_i^0\}, \\
& u_i =  \bra{\phi_i^0} H_{\textrm{cm}}^2 \ket{\phi_i^0} =  \bra{\phi_i^0} \bar{\bar{H}}_{\textrm{cm}}^2 \ket{\phi_i^0},\\
& \bar{x}_i = \sum_{m} \bra{\phi_m^0} x \ket{\phi_i^0}, \\
& \bar{y}_i = \sum_{m} \bra{\phi_m^0} y \ket{\phi_i^0}. 
\end{align}

Then the modified eigenstate is 
\begin{align}\label{state}
\ket{\phi_i} = \ket{\phi_i^0} + \ket{\phi_i^1} \simeq \exp\left[\frac{\gamma}{\lambda} \left(x + y -\bar{\bar{x}}_i -\bar{\bar{y}}_i \right) \right] \ket{\phi_i^0},
\end{align}
where $\bar{\bar{x}}_i$ and $\bar{\bar{y}}_i$ are given by
\begin{align}
\bar{\bar{x}}_i &=   \bar{x}_i  - \sum_{j \neq i}   \frac{\gamma (E_d^0 - E_h^0)   \left[\bra{\phi_i^0} x^2 \ket{\phi_j^0} - \bar{x}_i^{*} \bar{x}_j  \right]  }{\lambda (u_j - u_i)}, \\
\bar{\bar{y}}_i &=   \bar{y}_i  - \sum_{j \neq i}   \frac{\gamma (E_d^0 - E_h^0)   \left[\bra{\phi_i^0} y^2 \ket{\phi_j^0} - \bar{y}_i^{*} \bar{y}_j  \right]  }{\lambda (u_j - u_i)}. 
\end{align}
Equation (\ref{state}) shows that the bulk states can exponentially be localized in the non-Hermitian case, while they cannot if the perturbation Hamiltonian $H_{cm}^2$ is Hermitian (i.e., if $\gamma$ is pure imaginary).

\section{Bulk-state localization and winding number}
As explained in the main text, the winding number defined by the non-Hermitian Bloch Hamiltonian [see Eqs.~(1)-(5) in the main text] cannot correctly describe the bulk-corner correspondence in the second-order topological insulator. This deviation results from the non-Bloch-wave behavior of open-boundary eigenstates of a non-Hermitian Hamiltonian \cite{SMYaoarXiv:1804.04672, SMShunyuYao2018}, which leads to the bulk-state localization (see Fig.~3 in the main text, and a quantitative analysis in the previous section). In order to figure out this unexpected non-Bloch-wave behavior, and precisely characterize the topological invariants for non-Hermitian systems, modified complex wavevectors, rather than real ones, are proposed for calculating the winding number \cite{SMYaoarXiv:1804.04672, SMShunyuYao2018}. In this section, we discuss how to modify the topological index based on complex wavevectors. 

According to Eqs.~(\ref{wnEq1}) -- (\ref{phEq1}), in the basis $\Phi = (\Phi_{1, 1}, ~\Phi_{1, 2}, ..., ~\Phi_{L, L-1}, ~\Phi_{L, L})$, we solve the real-space eigenequation:
\begin{align}\label{wnEq2}
H_0 \phi = E \phi,
\end{align}
where the wavefunction $\phi$ is $\phi = (\varphi_{1, 1}, ~\varphi_{1, 2}, ~..., ~\varphi_{L, L-1}, ~\varphi_{L, L})^T$ with $\varphi_{n_x, n_y} = (\phi_{n_x, n_y, A}, ~\phi_{n_x, n_y, B}, \phi_{n_x, n_y, C}, ~\phi_{n_x, n_y, D})^T$. Then, according to Eq.~(\ref{wnEq2}), we have
\begin{align}\label{wnEq3}
T^{\dagger} \varphi_{n_x-1, n_y+1} + M^{\dagger} \varphi_{n_x, n_y} + R \varphi_{n_x, n_y+1} + M \varphi_{n_x, n_y+2} + T \varphi_{n_x+1, n_y+1} = E \varphi_{n_x, n_y+1},
\end{align}
where $R$, $M$ and $T$ are given by
\begin{align}\label{wnEq4}
& R = t (\tau_x + \tau_y \sigma_y) - i \gamma (\tau_y \sigma_z + \tau_y \sigma_x), \\
& M = \frac{\lambda}{2}  (\tau_y \sigma_y - i \tau_y \sigma_x), \\
& T = \frac{\lambda}{2}  (\tau_x + i \tau_y \sigma_z).
\end{align}

To derive the eigenequation [see Eq.~(\ref{wnEq3})], we consider the trial solution
\begin{align}\label{wnEq5}
\varphi_{n_x, n_y} = \exp(\alpha_1 n_x + \alpha_2 n_y) \phi_0,
\end{align} 
where $\phi_0= (\phi_{A}, ~\phi_{B}, ~\phi_{C}, ~\phi_{D})$. In order to preserve the symmetry $\mathcal{M}_{xy}$, we set $\alpha_1 = \alpha_2 = \alpha$. According to Eqs.~(\ref{wnEq3})--(\ref{wnEq5}), we have
\begin{align}\label{wnEq6}
(t - \gamma + \beta \lambda) \phi_C - (t -\gamma + \beta \lambda) \phi_D &  = E \phi_A, \\
\left[\lambda + \beta (t + \gamma)\right] \phi_C + \left[\lambda + \beta (t + \gamma)\right] \phi_D & = \beta E \phi_B, \\
\left[\lambda + \beta (t + \gamma)\right] \phi_A + \beta ( t -\gamma + \beta  \lambda) \phi_B & = \beta E \phi_C,  \\
-\left[\lambda + \beta (t + \gamma)\right] \phi_A + \beta ( t -\gamma + \beta  \lambda) \phi_B & = \beta E \phi_D, 
\end{align} 
where $\beta = \exp(\alpha)$. Therefore, we have
\begin{align}\label{wnEq7}
\lambda \gamma \beta^2 - \left (\gamma^2 - \lambda^2 \right) \beta + t^2 \beta - \gamma\lambda + \left(1 + \beta^2 \right)\lambda t = \frac{\beta}{2} E^2.
\end{align} 
Equation (\ref{wnEq7}) has two solutions $\beta_1$ and $\beta_2$:
\begin{align}\label{wnEq8}
\beta_i = \frac{2 \gamma^2-2 \lambda ^2 -2 t^2 + E^2 \pm \sqrt{\left(2 \lambda ^2 - 2 \gamma^2 + 2 t^2 - E^2\right)^2 - 8 \lambda^2 (  t - \gamma) ( \gamma + t)}}{4 \lambda(\gamma + t)}.
\end{align} 
Moreover, according to Eq.~(\ref{wnEq7}), for $E \to 0 $, we have
\begin{align}\label{wnEq9}
& \beta_1 = -\frac{\lambda}{t + \gamma} , ~~~\beta_2 = \frac{\gamma - t}{\lambda}, ~~~ t \in  [-\sqrt{\gamma^2 + \lambda^2}, ~~ \sqrt{\gamma^2 + \lambda^2}], \\ 
& \beta_1 = \frac{\gamma - t}{\lambda}, ~~~\beta_2 = -\frac{\lambda}{t + \gamma}, ~~~ t \in  [-\infty, ~~ -\sqrt{\gamma^2 + \lambda^2}] \cup [\sqrt{\gamma^2 + \lambda^2}, ~~ +\infty].
\end{align} 

Then, the state vector in Eq.~(\ref{wnEq5}) at each site can be written as
\begin{align}\label{wnEq10}
\varphi_{n_x, n_y} = \beta_1^{n_x + n_y} \phi_0^1 + \beta_2^{n_x + n_y} \phi_0^2.
\end{align} 

By considering the following boundary conditions
\begin{align}\label{wnEq11}
R (\beta_1^2 \phi_0^1 + \beta_2^2 \phi_0^2) + M (\beta_1^3 \phi_0^1 + \beta_2^3 \phi_0^2) + T (\beta_1^3 \phi_0^1+ \beta_2^3 \phi_0^2) &  = E (\beta_1^2 \phi_0^1 + \beta_2^2 \phi_0^2),\\
T^{\dagger} (\beta_1^{2 L - 1} \phi_0^1 + \beta_2^{2 L - 1} \phi_0^2) + M^{\dagger} (\beta_1^{2 L - 1} \phi_0^1 + \beta_2^{2 L - 1} \phi_0^2) + R (\beta_1^{2 L} \phi_0^1 + \beta_2^{2 L} \phi_0^2) & = E (\beta_1^{2 L} \phi_0^1 + \beta_2^{2 L} \phi_0^2),
\end{align} 
and relations
\begin{align}\label{wnEq12}
\phi_A^{(i)} & = \frac{\beta_i E (\phi_C^{(i)} - \phi_D^{(i)})}{2 (\lambda + t \beta + \gamma \beta)}, \\
\phi_B^{(i)} & = \frac{\beta_i E (\phi_C^{(i)} + \phi_D^{(i)})}{2 (t - \gamma + \lambda \beta)},
\end{align} 
we have
\begin{align}\label{wnEq13}
\beta _2^{2 L - 1} \left[2 t^2 -2 \gamma^2 + 2 \beta_1 \lambda  (\gamma + t) - E^2\right]= \beta_1^{2 L - 1} \left[2 t^2 -2 \gamma ^2 + 2 \beta_2 \lambda  (\gamma + t)-E^2\right].
\end{align}

According to Eq.~(\ref{wnEq13}), we require that $\beta_1$ and $\beta_2$ satisfy
\begin{align}\label{wnEq14}
\abs{\beta_1} = \abs{\beta_2}
\end{align}
for a continuum  spectrum, where the number of energy eigenstates is proportional to the lattice size $L$. Otherwise, $\beta_1(E) = 0 $ or $2 t^2 -2 \gamma^2 + 2 \beta_2 \lambda  (\gamma + t) - E^2 = 0$ (which is independent of the lattice size $L$) if $\abs{\beta_1} > \abs{\beta_2}$, and $\beta_2(E) = 0 $ or $2 t^2 -2 \gamma ^2 + 2 \beta_1 \lambda  (\gamma + t)-E^2$ (which is independent of the lattice size $L$) if $\abs{\beta_1} < \abs{\beta_2}$. 

By combining Eqs. (\ref{wnEq8}) and (\ref{wnEq14}), for the bulk states, we have
\begin{align}\label{wnEq15}
\beta_0 = \abs{\beta_i} = \sqrt{\abs{\frac{t-\gamma}{t+\gamma}}}.
\end{align}
Then, to account for the non-Bloch-wave behavior \cite{SMShunyuYao2018, SMYaoarXiv:1804.04672}, we replace the real vavevector $\textbf{k}$ with the complex one
\begin{align}\label{wnEq16}
\textbf{k} = (k_x, ~k_y) \, \to \,  \widetilde{\textbf{k}} =\textbf{k} + i \textbf{k}'= (k_x + i k_x', ~ k_y + i k_y'),
\end{align}
where 
\begin{align}\label{wnEq17}
k_x' = k_y' = - \textrm{ln}(\beta_0) = - \alpha_0.
\end{align}
Then the momentum-space Hamiltonian [see Eq.~(S2)] can be expressed as
\begin{align}\label{wnEq18}
H_{\textrm{2D}}(\mathbf{k}) \, \to \, \widetilde{H}(\textbf{k}) = H(\textbf{k} + i \textbf{k}').
\end{align}

As shown in the main text, because the Hamiltonian $\widetilde{H}(\textbf{k})$ preserves the mirror-rotation symmetry $\mathcal{M}_{xy}$ , we can write the Hamiltonian $\widetilde{H}(\textbf{k})$ in a block-diagonal form with $k_x = k_y = k$ as
\begin{align}\label{wnEq19}
U^{-1} \widetilde{H}(k,k) U = \left[\begin{matrix}
\widetilde{H}_{+}(k) & 0 \\
0 & \widetilde{H}_{-}(k) \\  \end{matrix}\right],
\end{align}
where $\widetilde{H}_{+}(k)$ acts on the $+1$ mirror-rotation subspace, and $\widetilde{H}_{-}(k)$ acts on the $-1$ mirror-rotation subspace. The the unitary transformation $ U $ is
\begin{align}
U = \left[\begin{matrix}
0 & 0 & 1 & 0 \\
1 & 0 & 0 & 0 \\
0 & \frac{1}{\sqrt{2}} & 0 & \frac{1}{\sqrt{2}} \\
0 & \frac{1}{\sqrt{2}} & 0 & -\frac{1}{\sqrt{2}}  \\  \end{matrix}\right],
\end{align}
and $\widetilde{H}_{+}(k)$ and $\widetilde{H}_{-}(k)$ are
\begin{align}\label{wnEq20}
\widetilde{H}_{+}(k) = \sqrt{2} \left[t + \lambda \cos(k - i \alpha_0) \right] \sigma_x + \sqrt{2} \left[\lambda \sin(k - i \alpha_0) + i \gamma \right] \sigma_y, 
\end{align}
\begin{align}\label{wnEq21}
\widetilde{H}_{-}(k) = \sqrt{2} \left[t + \lambda \cos(k - i \alpha_0) \right] \sigma_x - \sqrt{2} \left[\lambda \sin(k - i \alpha_0) + i \gamma \right] \sigma_y. 
\end{align}
We rewrite Eqs.~(\ref{wnEq20})--(\ref{wnEq21}) as 
\begin{align}\label{wnEq22}
\widetilde{H}_{+}(k) = \sqrt{2} \left(t + \gamma + \lambda \beta_0^{-1} \textrm{e}^{-i k} \right) \sigma_{+} + \sqrt{2} \left(t - \gamma + \lambda \beta_0 \textrm{e}^{i k} \right) \sigma_{-}, 
\end{align}
\begin{align}\label{wnEq23}
\widetilde{H}_{-}(k) = \sqrt{2} \left(t - \gamma + \lambda \beta_0 \textrm{e}^{i k} \right) \sigma_{+} + \sqrt{2} \left(t + \gamma + \lambda \beta_0^{-1} \textrm{e}^{-i k} \right) \sigma_{-},
\end{align}
where $\sigma_{\pm} = (\sigma_x \pm i \sigma_y)/2$. The winding numbers for the Hamiltonians $\widetilde{H}_{+}(k)$ and $\widetilde{H}_{-}(k)$ are expressed as

\begin{align}\label{wnEq24}
w_{+} = \frac{i}{2 \pi} \int_{0}^{4 \pi} \! \frac{\bra{\chi_{+}} \partial_k \ket{\phi_{+}}}{\bra{\chi_{+}} \ket{\phi_{+}}} dk,  
\end{align}
\begin{align}\label{wnEq25}
w_{-} = \frac{i}{2 \pi} \int_{0}^{4 \pi} \! \frac{\bra{\chi_{-}} \partial_k \ket{\phi_{-}}}{\bra{\chi_{-}} \ket{\phi_{-}}} dk,  
\end{align}                                                                                      %
where $\ket{\phi_{\pm}}$ and $\ket{\chi_{\pm}}$ are the right and left eigenstates of $\widetilde{H}_{\pm}(k)$, respectively. The integration over $4 \pi$ in Eqs.~(\ref{wnEq24})--(\ref{wnEq25}) is attributed to the $4 \pi$-periodicity of the eigenenergies and eigenstates. Then, the total winding number is 
\begin{align}\label{wnEq26}
w = w_{+} - w_{-}.  
\end{align}

\section{Possible experimental realization}
The second-order topological insulator studied here can be experimentally realized in ultracold atoms and photonic systems. In this section, we propose a possible scheme to realize a non-Hermitian second-order topological insulator in ultracold atoms in optical lattices. The idea is to combine two state-of-the-art experimental techniques: artificial gauge field and dissipation engineering. The former is used to create a $\pi$ flux in each unit cell and the latter is used to make the hopping amplitudes asymmetric.

\subsection{General idea}        
We note that the anti-Hermitian part of the Hamiltonian can be separated into individual four-site blocks [see Eq.~(S1)]. By individual we mean that the corresponding local Hamiltonians have no overlap with each other. To be specific, we consider the following four-site Hamiltonian (here we omit the notation $R$ in operators for the sake of simplicity):
\begin{equation}
H_{\rm sb}=(t+\gamma)[c^\dag_2(c_3+c_4)+(c^\dag_3-c^\dag_4)c_1]+(t-\gamma)[(c^\dag_3+c^\dag_4)c_2+c^\dag_1(c_3-c_4)],
\label{Hsb}
\end{equation}
whose Hermitian part reads
\begin{equation}
H_0=\frac{1}{2}(H_{\rm sb}+H^\dag_{\rm sb})=t[c^\dag_2(c_3+c_4)+(c^\dag_3-c^\dag_4)c_1+(c^\dag_3+c^\dag_4)c_2+c^\dag_1(c_3-c_4)].
\end{equation}
On the other hand, the anti-Hermitian part of Eq.~(\ref{Hsb}) is given by
\begin{equation}
\frac{1}{2}(H_{\rm sb}-H^\dag_{\rm sb})=\gamma(c^\dag_3c_1+c^\dag_1c_4+c^\dag_2c_3+c^\dag_2c_4-{\rm H.c.}).
\label{nonHsb}
\end{equation}
To engineer this anti-Hermitian part, 
we follow the method developed in Ref.~\cite{SMarXiv:1802.07964} to use a combination of the following jump operators that describe the collective loss of two nearest-neighbor sites:
\begin{equation}
L_1=\sqrt{2\gamma}(c_3+ic_1),\;\;\;\;
L_2=\sqrt{2\gamma}(c_1+ic_4),\;\;\;\;
L_3=\sqrt{2\gamma}(c_2+ic_3),\;\;\;\;
L_4=\sqrt{2\gamma}(c_2+ic_4).
\label{JPO}
\end{equation}
At the single-particle or mean-field level, the open-system dynamics of a single block is determined by the non-Hermitian effective Hamiltonian
\begin{equation}
H_{\rm eff}=H_0-\frac{1}{2}\sum^4_{j=1}L^\dag_jL_j=H_{\rm sb}-2\gamma\sum^4_{j=1}c^\dag_jc_j,
\label{Heff}
\end{equation}
which differs from Eq.~(\ref{Hsb}) only by a background loss term proportional to $2\gamma$. In the following, we will discuss in detail a possible implementation of the above idea with state-of-the-art experimental techniques developed in dissipation engineering \cite{SMMULLER20121} and artificial gauge fields \cite{SMGoldman2014}.

\subsection{Explicit implementation}
We first note that, in the Hermitian limit ($\gamma=0$), there is already an ultracold-atom-based proposal in Ref.~\cite{SMBenalcazar61}. As explained therein, the Hermitian Hamiltonian can be simulated by embedding a superlattice structure into a two-dimensional $\pi$-flux lattice, which can be realized using a setup described in Refs.~\cite{SMBloch2013,SMKetterle2013}. Moreover, we would like to mention that a sharp (box) boundary is also available within current experimental techniques \cite{SMGaunt2013}. This is necessary for observing the topologically protected corner states.

\begin{figure}
	\begin{center}
		\includegraphics[clip,width=17cm]{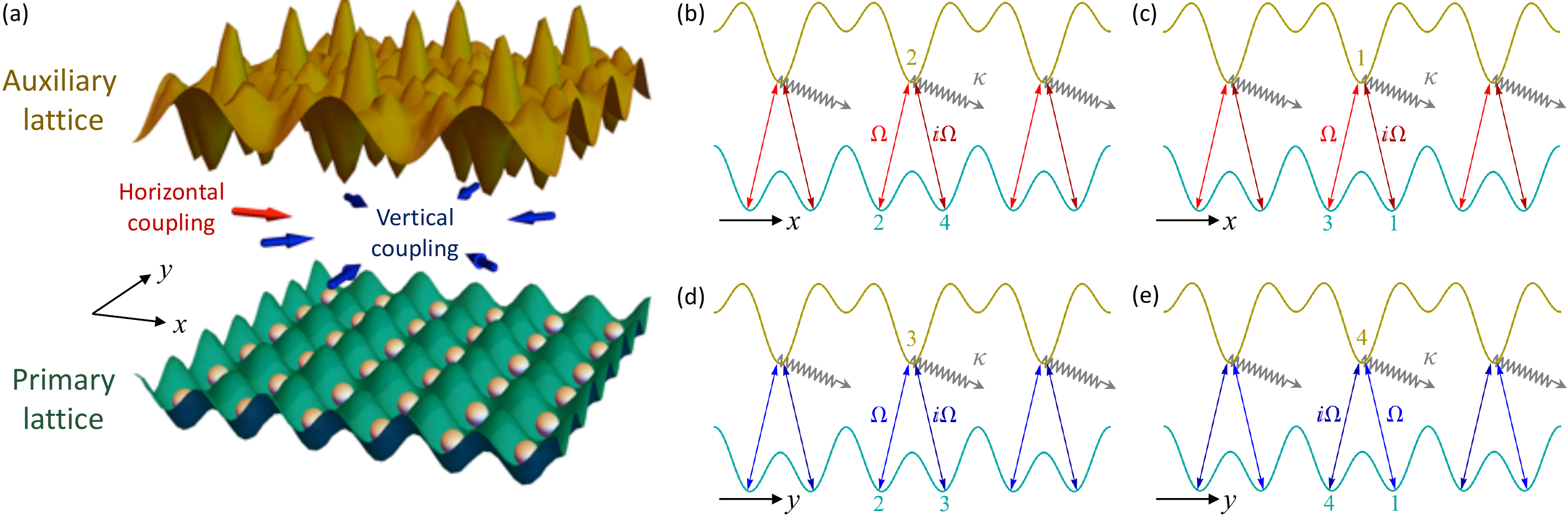}
	\end{center}
	\caption{(a) Schematic illustration of a proposed experimental setup. The primary lattice (green) together with a pair of Raman lasers (not shown) gives rise to a Hermitian second-order topological insulator, where the Raman lasers are used for inducing effective particle hopping. The asymmetry in hopping amplitudes is then introduced via a coherent coupling to a dissipative auxiliary lattice (yellow). In particular, the running wave (red arrow) generates the horizontal couplings shown in (b) and (c), which are the cross-sections along the $x$ direction, containing sublattices $2$, $4$ and $3$, $1$, respectively. The other three standing waves (blue arrows) generate the vertical couplings shown in (d) and (e), which are the cross-sections along the $y$ direction containing sublattices $2$, $3$ and $4$, $1$, respectively.}
	\label{FigSM6}
\end{figure}

Now let us move onto the asymmetric hopping amplitudes. Following the theoretical consideration sketched out above, it suffices to focus on the realization of the jump operators [Eq.~(\ref{JPO})]. We again follow in Ref.~\cite{SMarXiv:1802.07964} to effectively engineer a collective loss from a combination of on-site loss of auxiliary states and their coherent coupling to the primary degrees of freedom. 
Without loss of generality, we now focus on the case of a single block and resonant couplings. As shown in Fig.~\ref{FigSM6}, the full open-system dynamics in the rotating frame of reference can be written as
\begin{equation}
\dot\rho_t=-i[H_0+\frac{\Omega}{2}(a^\dag_2(c_2+ic_4)+a^\dag_1(c_3+ic_1)+a^\dag_3(c_2+ic_3)+a^\dag_4(c_1+ic_4)+{\rm H.c.}),\rho_t]+\kappa\sum^4_{j=1}\mathcal{D}[a_j]\rho_t,
\label{effdyn}
\end{equation}
where $a_j$'s denote the annihilation operators of the particles in the auxiliary sublattice $j$ and $\mathcal{D}[L]\rho\equiv L\rho L^\dag-\frac{1}{2}\{L^\dag L,\rho\}$ is the Lindblad superoperator. In the regime $\kappa\gg\Omega$, we can adiabatically eliminate the fast decay modes in the auxiliary lattice \cite{SMReiter2012} to obtain the following effective dynamics of the primary lattice degrees of freedom alone:
\begin{equation}
\dot\rho_t=-i[H_0,\rho_t]+2\gamma\sum^4_{j=1}\mathcal{D}[L_j]\rho_t,
\end{equation}
where $\gamma=\Omega^2/(2\kappa)$. 
At the single-particle or the mean-field level, the dynamics reduces to the nonunitary evolution governed by the non-Hermitian Hamiltonian given in Eq.~(\ref{Heff}).

However, unlike the simple case discussed in Ref.~\cite{SMarXiv:1802.07964}, which describes a single band in one dimension, here we have two difficulties to realize the effective dynamics in Eq.~(\ref{effdyn}): (i) To ensure that an auxiliary site is only coupled to two nearest-neighbor sites, the auxiliary lattice should form a square-octagon pattern in two dimensions; (ii) We have to fine-tune the phases of couplings in the presence of a nonzero flux. Let us discuss below possible solutions to (i) and (ii).

\begin{figure}
	\begin{center}
		\includegraphics[clip,width=17cm]{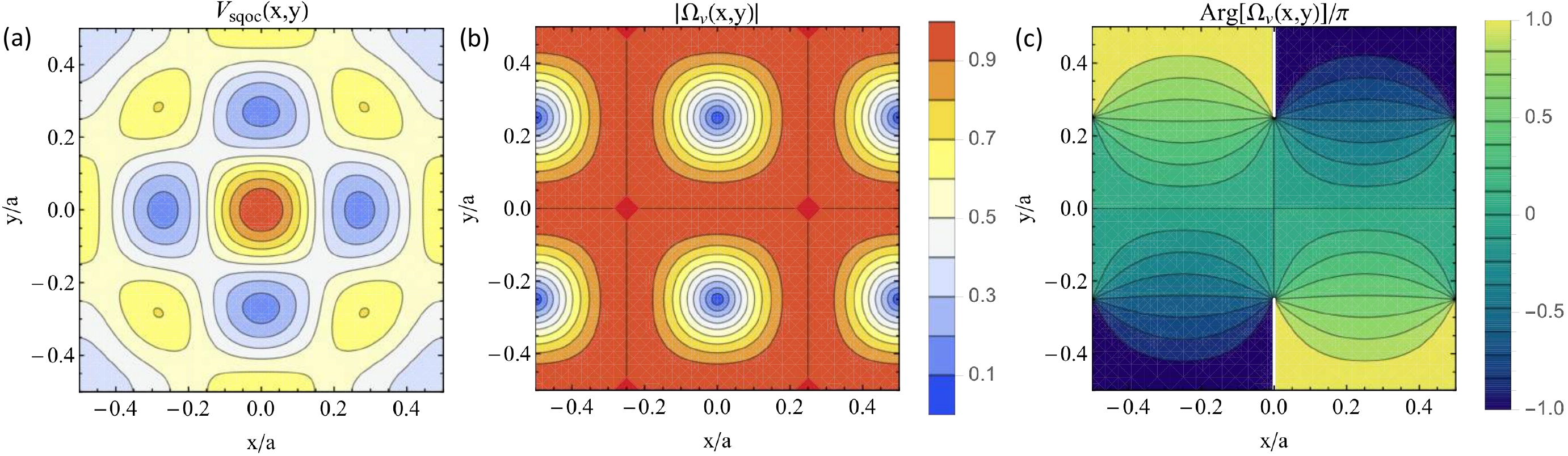}
	\end{center}
	\caption{(a) Square-octagon pattern of the auxiliary lattice potential $V_{\rm sqoc}(\boldsymbol{r})$ [see also Fig.~\ref{FigSM6}(a)]. (b) Magnitude and (c) phase patterns of the vertical Rabi coupling $\Omega_{\rm v}(\boldsymbol{r})$ given in Eq.~(\ref{Rabiv}). The units in (a) and (b) are set to be the largest $V_{\rm sqoc}(\boldsymbol{r})$ and $|\Omega_{\rm v}(\boldsymbol{r})|$.}
	\label{FigSM7}
\end{figure}

To overcome the difficulty (i), we first note that an ideal trap with square-octagon geometry is given by 
\begin{equation}
V_{\rm idsqoc}(\boldsymbol{r})\propto\sum_{\substack{(m,n)\in\mathbb{Z}^2,\\s=\pm, \varsigma=0,1}}\delta \! \left(x-\left[\frac{s\varsigma}{4}+m \right]a\right) \delta \! \left(y-\left[\frac{1-\varsigma}{4}s+n\right]a\right),
\end{equation}
which can be expanded into the Fourier series 
\begin{equation}
V_{\rm idsqoc}(\boldsymbol{r})\propto\sum_{(m,n)\in\mathbb{Z}^2}\left[i^m+(-i)^m+i^n+(-i)^n\right] \exp(i\frac{2\pi}{a}[mx+ny]).
\label{VidFS}
\end{equation}
Here $a$ is the lattice constant, which equals $\lambda_{\rm l}$, the wavelength of the laser that generates the primary lattice \cite{SMBenalcazar61}. Keeping the terms with $|m|+|n|=3,4$ in Eq.~(\ref{VidFS}) followed by dropping the constants and adjusting the coefficients, we can construct a square-octagon-lattice potential as
\begin{equation}
V_{\rm sqoc}\propto \left[\cos^2\frac{\pi(2x+y)}{a}+\cos^2\frac{\pi(2x-y)}{a}+\cos^2\frac{\pi(x+2y)}{a}+\cos^2\frac{\pi(x-2y)}{a}+\cos^2\frac{2\pi(x-y)}{a}+\cos^2\frac{2\pi(x+y)}{a}\right],
\end{equation}
whose profile in a single unit cell is plotted in Fig.~\ref{FigSM7}(a). In practice, this potential can be generated by six standing-wave lasers with amplitude profiles given by
\begin{equation}
\cos\frac{2\pi[\cos\theta(2x\pm y)/\sqrt{3}+\sin\theta z]}{\lambda_{\rm al}}, ~ \cos\frac{2\pi[\cos\theta(x\pm 2y)/\sqrt{3}+\sin\theta z]}{\lambda_{\rm al}}, ~\textrm{and}~ \cos\frac{2\pi[\cos\theta'(x\pm y)/\sqrt{2}+\sin\theta' z]}{\lambda_{\rm al}}, 
\end{equation}
where $\theta=\arccos(\sqrt{3}\lambda_{\rm al}/2a)$ and $\theta'=\arccos(\sqrt{2} \lambda_{\rm al} /a)$ are the tilt angles from the $x$-$y$ plane. Therefore, we can rather freely choose $\lambda_{\rm al}$ such that the auxiliary lattice only selectively traps a certain metastable state of atoms, such as the ${}^3$P${}_0$ state of alkaline-earth atoms. The on-site loss rate $\kappa$ can be controlled by the strength of an additional laser that couples the metastable state to a certain unstable state, such as the ${}^1$P${}_1$ state of alkaline-earth atoms.

To overcome the difficulty (ii), we first note that the effective Rabi coupling $\Omega_{\rm tb}$ within the tight-bonding approximation can be related to the spatial distribution of the Rabi coupling $\Omega(\boldsymbol{r})$ via
\begin{equation}
\Omega_{\rm tb}=\int d^2\boldsymbol{r} \; \Omega(\boldsymbol{r})         
\; w^*_{\rm a}(\boldsymbol{r}-\boldsymbol{r}_{\rm a}) \; w_{\rm l}(\boldsymbol{r}-\boldsymbol{r}_{\rm l}),
\end{equation}
where $\omega_{\rm l}$ and $\omega_{\rm a}$ are the Wannier functions of the primary and auxiliary lattices, respectively. Therefore, the horizontal couplings $a^\dag_2(c_2+ic_4)$ and $a^\dag_1(c_3+ic_1)$ can easily be achieved by a running wave $\Omega_{\rm h}(\boldsymbol{r})\propto \exp(i\frac{2\pi}{a}x)$ along the $x$ direction, with the phase difference between $c_2$ ($c_3$) and $c_4$ ($c_1$) imprinted by the spatial phase variation of $\Omega_{\rm h}(\boldsymbol{r})$. On the other hand, for the vertical couplings $a^\dag_3(c_2+ic_3)$ and $a^\dag_4(c_1+ic_4)$, we cannot simply apply a running wave along the $y$ direction; otherwise we will obtain $a^\dag_4(c_4+ic_1)$ instead of $a^\dag_4(c_1+ic_4)$. To imprint the desired phase information, we can engineer the Rabi-frequency pattern to be
\begin{equation}
\Omega_{\rm v}(\boldsymbol{r})\propto \cos\frac{2\pi y}{a}-i\sin\frac{2\pi x}{a}\sin\frac{2\pi y}{a},
\label{Rabiv}
\end{equation}
which can be created by three standing wave lasers with amplitude profiles 
\begin{equation}
\cos\frac{2\pi(\cos\alpha y+\sin\alpha z)}{\lambda_{\rm c}} ~\textrm{and}~ \cos\frac{2\pi[\cos\alpha' (x\pm y)/\sqrt{2}+\sin\alpha' z]}{\lambda_{\rm c}}, 
\end{equation}
where $\alpha=\arccos(\lambda_{\rm c}/a)$ and $\alpha'=\arccos(\sqrt{2}\lambda_{\rm c}/a)$ are the tilt angles from the $x$-$y$ plane. We plot the spatial pattern of the magnitude and phase parts of $\Omega_{\rm v}$ [Eq.~(\ref{Rabiv})] in Figs.~\ref{FigSM7}(b) and (c), where we can clearly find some vortices indicated by the zeros of $|\Omega_{\rm v}(\boldsymbol{r})|$ accompanied by phase windings. This observation is consistent with the existence of a $\pi$ flux in the primary lattice.

\section{Non-Hermitian second-order topological phases in 3D}

\subsection{Model} 
We consider a minimal model of a 3D non-Hermitian SOTI on a cubic lattice:
\begin{align}\label{3DHamiltonianEq1}
H_\textrm{3D}(\mathbf{k})  = & \left[m + t \left(\cos k_x + \cos k_y + \cos k_z\right)\right] \tau_z + \left[\left(\Delta_1 \sin k_x + i \gamma_0 \right) \sigma_x + \left(\Delta_1 \sin k_y + i \gamma_0 \right) \sigma_y + \left(\Delta_1 \sin k_z + i \gamma_z \right)  \sigma_z \right] \tau_x \nonumber \\
& + \Delta_2 \left(\cos k_x - \cos k_y \right) \tau_y,
\end{align}
where we have set the lattice constant $a_0 = 1$, $\sigma_i$ and $\tau_i$ for $i = x, y, z$ are Pauli matrices acting on spin and orbital/sublattice degrees of freedom, respectively, and $m, t, \Delta_1, \Delta_2$, $\gamma_0$ and $\gamma_z$  are real parameters. Note that the Hermitian part of $H_\textrm{3D}(\mathbf{k})$ supports chiral hinge modes propagating along the $z$ direction \cite{SMTitusSciAdv2018}.


For $\Delta_2 = 0$, the Hermitian part of $H_\textrm{3D}(\mathbf{k})$ is invariant under time reversal $\mathcal{T} = \sigma_y \mathcal{K}$, where $\mathcal{K}$ being the complex conjugation operator, $\mathcal{M}_x = i \sigma_x \tau_z $ is the \textit{x} mirror reflection , $\mathcal{M}_y = i\sigma_y \tau_z $ is the \textit{y} mirror reflection, and $C_4 = \exp(-i \pi \sigma_z /4)$ is the $\pi/2$ rotation about the $z$ axis. These symmetries are broken by both non-Hermitian terms including $\gamma_0$ and terms including $\Delta_2$. However, $H_\textrm{3D}(\mathbf{k})$ is invariant under the mirror-rotation symmetry operation $\mathcal{M}_{xy} = C_4 \mathcal{M}_y$ with
\begin{align}\label{3DsymmetryEq1}
\mathcal{M}_{xy} H_\textrm{3D}(k_x, k_y, k_z) \mathcal{M}_{xy}^{-1} = H_\textrm{3D}(k_y, k_x, k_z). 
\end{align}

The bulk energy bands of the Hamiltonian $H_\textrm{3D}(\mathbf{k})$ are obtained as
\begin{align}\label{3DenegyEq1}
\bar{E}_{\pm}(\mathbf{k}) = & \pm \left[ -2 \gamma _0^2- \gamma _z^2 +\left(t \cos k_x +t \cos k_y + t \cos k_z + m\right){}^2+2 i  \Delta _1 \left(\gamma _0 \sin k_x + \gamma _0 \sin k_y + \gamma _z \sin k_z \right) \right. \nonumber \\
& \left. + ~ \Delta _2^2 \left(\cos k_x - \cos k_y \right){}^2 - \Delta _1^2 \left(\cos 2 k_x +\cos 2 k_y + \cos 2 k_z - 3\right)/2 \right]^\frac{1}{2}.
\end{align}
where the upper and lower branches are two-fold degenerate, respectively. The Hamiltonian $H_\textrm{3D}(\mathbf{k})$ is defective at the exceptional points (EPs) with
\begin{align}\label{3DenegyEq2}
\bar{E}_{\pm} (\mathbf{k}_\textrm{EP}) = 0,
\end{align}
if one of the following conditions is satisfied:

(1) EPs appear when $k_x = 0$ and $k_y = 0$, and then Eq.~(\ref{3DenegyEq2}) reduces to
\begin{align}\label{3DenegyEq3}
(t \cos k_z + m + 2 t)^2 - \left(\gamma_z - i \Delta _1 \sin k_z\right)^2 = 2 \gamma _0^2.
\end{align}

(2) EPs appear when $k_x = \pi$ and $k_y = \pi$, and then Eq.~(\ref{3DenegyEq2}) reduces to
\begin{align}\label{3DenegyEq4}
(t \cos k_z + m-2 t)^2 - \left(\gamma_z - i \Delta _1 \sin k_z\right)^2 = 2 \gamma _0^2.
\end{align}

(3) EPs appear when $k_x = 0 ~(\pi)$ and $k_y = \pi ~(0)$, and then Eq.~(\ref{3DenegyEq2}) reduces to
\begin{align}\label{3DenegyEq5}
4 \Delta _2^2 + (t \cos k_z + m)^2 - \left(\gamma_z - i \Delta _1 \sin k_z\right)^2   = 2 \gamma _0^2.
\end{align}

(4) EPs appear when $k_x = -k_y$ with $k_y \neq 0$ and $k_y \neq \pi$, and then Eq.~(\ref{3DenegyEq2}) reduces to
\begin{align}\label{3DenegyEq6}
(2 t \cos k_y + t \cos k_z + m)^2 -\Delta _1^2 (\cos 2 k_y + \cos 2 k_z/2-3/2)  - \gamma_z^2 + 2 i \gamma_z \Delta_1 \sin k_z  = 2 \gamma _0^2.
\end{align}
\begin{figure}[!tb]
	\centering
	\includegraphics[width=10cm]{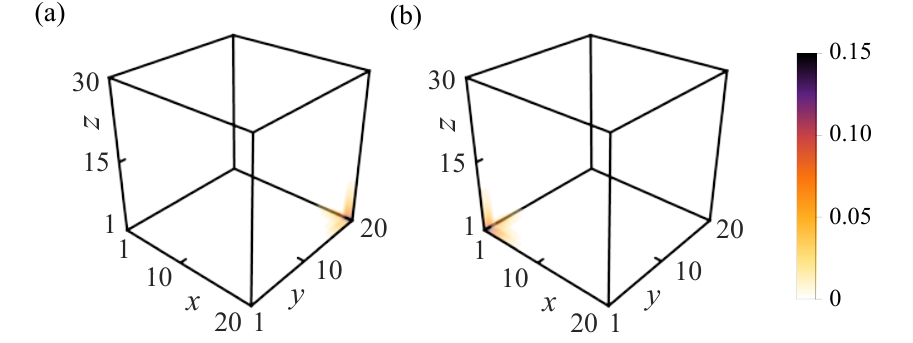}
	\caption{Probability density distributions $\abs{\Phi_{n, R}}^2$ ($n$ is the index of an eigenstate and $R$ specifies a lattice site) of mid-gap modes for a 3D non-Hermitian SOTI with open boundaries along all the directions (a) for $m = -2$ and $\gamma_z = -0.2$, and (b) for $m = 2$ and $\gamma_z = 0.2$. The mid-gap states (with eigenenergy of 0.035) are only localized at one corner on the $x = y$ plane. The number of unit cells is $20 \times 20 \times 30$ with $t = 1$, $\gamma_0 = 0.7$, $\Delta_1 = 1.2$, and $\Delta_2 = 1.2$.}\label{figSM10}
\end{figure}

\subsection{B. Localization of second-order boundary states at one corner} 
As shown in the main text, the mid-gap states in the 3D non-Hermitian SOTI are localized at the right corner of the $x = y$ plane [see Fig.~5(e) in the main text and Fig.~\ref{figSM10}(a)]. We note that the mid-gap states can also be localized at the left corner on the $x=y$ plane when an opposite sign of the parameter $m$ is considered, as shown in Fig.~\ref{figSM10}(b). The localization of the mid-gap states at one corner on the $x = y$ plane results from the interplay between mirror-rotation symmetry $\mathcal{M}_{xy}$ and non-Hermiticity.

\subsection{Effect of a different type of asymmetric hopping and non-Hermiticity on the localization of boundary states}

\begin{figure}[!tb]
	\centering
	\includegraphics[width=18cm]{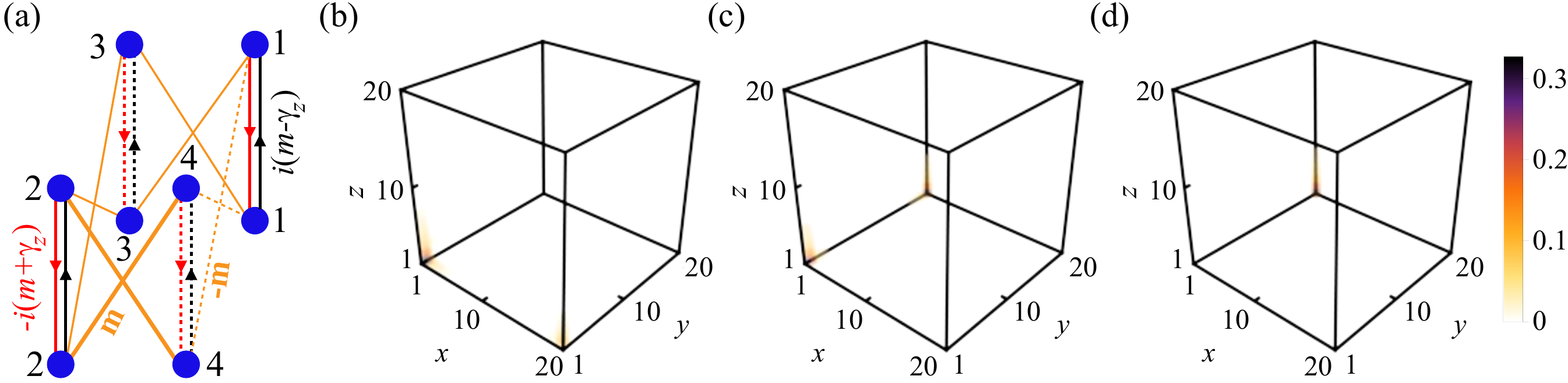}
	\caption{(a) Real-space representation of the 3D model in Eq.~(\ref{Nonreciprocal1}), which shows the particle hopping along the $z$ direction between 2D layers. This 3D lattice is constructed by stacking the 2D lattice layer shown in Fig.~\ref{figSM15}(a). Along the $z$ direction, the nearest and next-to-nearest neighbor hoppings are considered. The dashed lines indicate hopping terms with a negative sign.  Probability density distributions $\abs{\Phi_{n, R}}^2$ ($n$ is the index of an eigenstate and $R$ specifies a lattice site) of mid-gap modes with open boundaries along all the directions: (b) for  $\gamma_x = 0$ and $\gamma_y = 0.5$, (c) for $\gamma_x = 0.5$ and $\gamma_y = 0$, and (d) for $\gamma_x = 0.5$ and $\gamma_y = -0.5$. The mid-gap states (with eigenenergy of 0.04) are localized at more than one corner for $\gamma_x = 0$ or $\gamma_y = 0$. The number of unit cells is 20 $\times$ 20 $\times$ 20 with $\gamma_z = -0.2$, $m = -1$, $t_x = t_y = 1.0$ and $\lambda = 1.5$.}\label{figSM25}
\end{figure}
\begin{figure}[!tb]
	\centering
	\includegraphics[width=18cm]{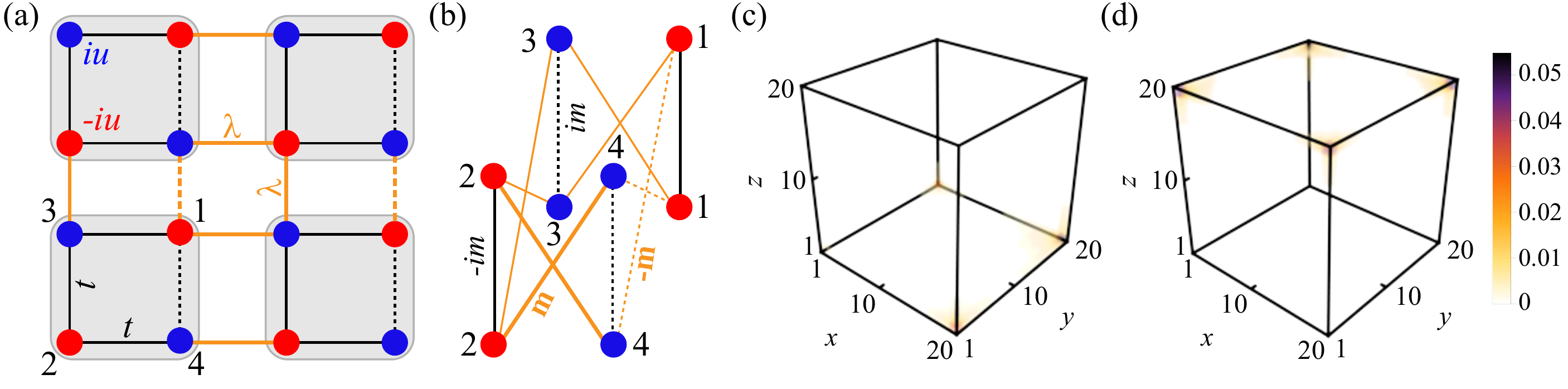}
	\caption{(a,b) Real-space representation of the 3D model in Eq.~(\ref{Nonreciprocal2}). The particle hopping in the $(x, y)$-plane, with the alternating on-site gain and loss (with the imaginary staggered potentials $iu$ and $-iu$) within each unit cell, is shown in (a). The particle hopping along the $z$ direction between 2D layers is indicated in (b). This 3D lattice is constructed by stacking the 2D lattice layer shown in (a). The dashed lines indicate hopping terms with a negative sign.  Probability density distributions $\abs{\Phi_{n, R}}^2$ ($n$ is the index of an eigenstate and $R$ specifies a lattice site) of mid-gap modes with open boundaries along all the directions: (c) for  $u = -0.2$, and (d) for $u = 0.2$. The mid-gap states (with eigenenergy of 0.04) are localized at more than one corner. The number of unit cells is 20 $\times$ 20 $\times$ 20 with $m = -1$, $t = 1.0$ and $\lambda = 1.5$.}\label{figSM35}
\end{figure}

In the 2D model, we have considered the effect of a different type of asymmetric hopping and non-Hermiticity (i.e., balanced gain and loss) on the localization of corner states in Sec.~V. In this section, we study effect of a different type of asymmetric hopping and non-Hermiticity on the localization of second-order boundary modes in 3D  non-Hermitian SOTIs. The non-Hermitian terms of the 3D model studied in our manuscript take the form of an imaginary Zeeman field \cite{SMYaoarXiv:1804.04672, SMPhysRev.87.410}. Therefore, to investigate the effect of a different type of asymmetric hopping on the localization of second-order boundary modes in 3D SOTIs, we consider a 3D non-reciprocal lattice model by stacking the 2D lattice layers as shown in Fig.~\ref{figSM15}(a). The particle hoppings along the $z$ direction between 2D layers are indicated in Fig.~\ref{figSM25}(a). In momentum space, the Hamiltonian has the form
\begin{align}\label{Nonreciprocal1}
\bar{H}_{\textrm{3D}}(\mathbf{k}) & = \left[t_x + \lambda \cos(k_x) \right] \tau_x - \left[\lambda \sin(k_x) ~ + i \gamma_x \right] \tau_y \sigma_z +  \left[t_y + \lambda \cos(k_y) \right] \tau_y \sigma_y + \left[\lambda \sin(k_y) + i \gamma_y \right] \tau_y \sigma_x  \nonumber \\
&~~~~ + m \cos(k_z) (\tau_x + \tau_y \sigma_y) + \left[m \sin(k_z) - i \gamma_z \cos(k_z) \right] \tau_z,
\end{align}
where the Hermitian part of the Hamiltonian $\bar{H}_{\textrm{3D}}(\mathbf{k})$ supports four-degenerate chiral hinge modes propagating along the $z$ direction. Note that 3D Hermitian SOTIs can be present even in the absence of crystalline symmetries, as in the 2D case \cite{SMPhysRevLett.119.246401}.

When the larger hopping strength with amplitude $-i(m+\gamma_z)$ (i.e., $m$ and $\gamma_z$ have the same signs) along the $z$ direction is considered  [see Fig.~\ref{figSM25}(a)], the second-order boundary modes are localized at corners of the bottom side (i.e., $z=1$ plane), as shown in Figs.~\ref{figSM25}(b-d). In this case, for the larger hopping amplitude with $(t_y + \gamma_y)$ along the $y$ direction and the symmetric hopping along the $x$ direction, the second-order boundary states are localized at both the lower-left and lower-right corners of $z=1$ plane  [see Fig.~\ref{figSM25}(b)]. In contrast, these second-order boundary modes are localized at both the lower-left and upper-left corners for the larger hopping amplitude with $(t_x + \gamma_x)$ along the $x$ direction and the symmetric hopping along the $y$ direction [see Fig.~\ref{figSM25}(c)]. Moreover, the asymmetric hoppings along both the $x$ and $y$ directions make the second-order boundary modes localized at one corner [see Fig.~\ref{figSM25}(d)]. Therefore, the localization of the second-order boundary modes in the 3D non-reciprocal lattice model also depends on the type of asymmetric hopping. 

In addition to the different type of asymmetric hopping, the second-order boundary modes in 3D systems can be localized at more than one corner in the presence of a different type of non-Hermiticity i.e., balanced gain and loss. As shown in Figs.~\ref{figSM35}(a) and (b), we consider the alternating on-site gain and loss with symmetric particle hopping, where the imaginary staggered potentials are indicated by $iu$ and $-iu$. The Bloch Hamiltonian is written as
\begin{align}\label{Nonreciprocal2}
\tilde{H}_{\textrm{3D}}(\mathbf{k}) & = \left[t + \lambda \cos(k_x) \right] \tau_x - \lambda \sin(k_x)  \tau_y \sigma_z  +  \left[t + \lambda \cos(k_y) \right] \tau_y \sigma_y + \lambda \sin(k_y)  \tau_y \sigma_x  \nonumber \\
&~~~~ + m \cos(k_z) (\tau_x + \tau_y \sigma_y) + m \sin(k_z) \tau_z - i u \tau_z.
\end{align}

Figures \ref{figSM35}(c) and (d) show the probability density distributions for  $u = -0.2$ and $u = 0.2$, respectively. In the presence of the balanced gain and loss, the second-order boundary modes are localized at more than one corner of either the top or the bottom side in 3D systems.

\subsection{Low-energy effective Hamiltonians} 
\begin{figure}[!tb]
	\centering
	\includegraphics[width=8cm]{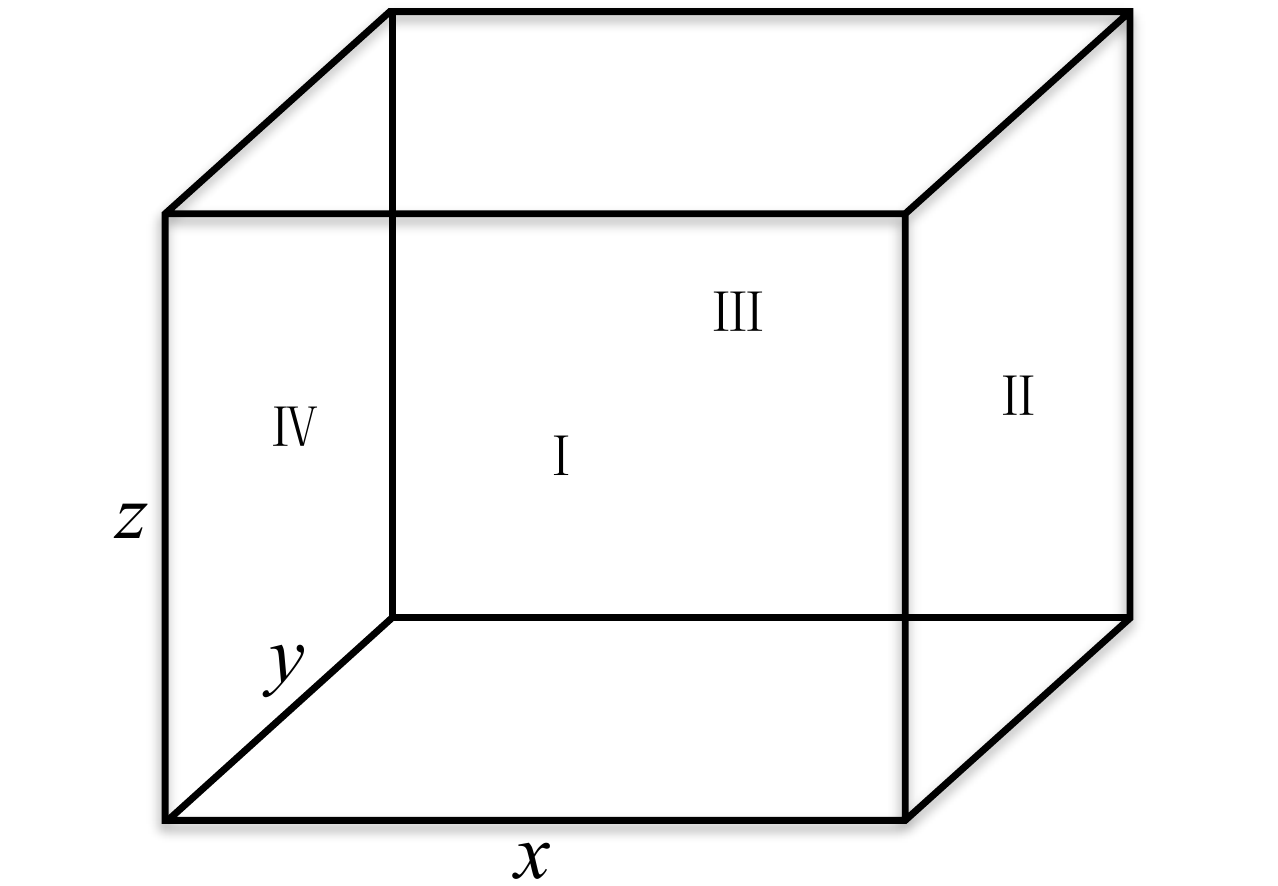}
	\caption{Schematic illustration of a 3D non-Hermitian SOTI in a cubic sample. $\textrm{\Rmnum{1}}$, $\textrm{\Rmnum{2}}$, $\textrm{\Rmnum{3}}$ and $\textrm{\Rmnum{4}}$ label the four surfaces of the lattice.}\label{figSM11}
\end{figure}

The localization of the mid-gap states at one corner on the $x=y$ plane results from  the symmetry $\mathcal{M}_{xy}$ and non-Hermiticity, and each mid-gap mode is a mutual topological state of two intersecting surfaces parallel to the $z$-axis. In this section, we present low-energy effective Hamiltonians to explain this effect. We label the four surfaces of a cubic sample as $\textrm{\Rmnum{1}}, \textrm{\Rmnum{2}}, \textrm{\Rmnum{3}}, ~\textrm{and}~ \textrm{\Rmnum{4}}$ (see Fig.~\ref{figSM11}). For the sake of simplicity, we consider the case of $\min\{-m, t, \Delta_1, \Delta_2\} > 0$. In this case, the low-energy bands of the Hermitian part of $H_{\textrm{3D}}(\mathbf{k})$ lie around the $\Gamma$ point of the Brillouin zone. Therefore, we consider a continuum model of the lattice Hamiltonian [see Eq.~(\ref{3DHamiltonianEq1})] by expanding its wavevector $\mathbf{k}$ to second order around the $\Gamma = (0, 0, 0)$ point of the Brillouin zone, obtaining
\begin{align}\label{EffHEq1}
\bar{H}_\textrm{cm}(\mathbf{k})  =  & \left[m + 3 t -\frac{t}{2}\left(k_x^2 + k_y^2 + k_z^2\right)\right] \tau_z + \left[\left(\Delta_1 k_x + i \gamma_0 \right) \sigma_x + \left(\Delta_1  k_y + i \gamma_0 \right) \sigma_y + \left(\Delta_1 k_z+ i \gamma_z \right) \sigma_z \right] \tau_x \nonumber \\
& + \frac{\Delta_2}{2} \left(k_y^2 -  k_x^2 \right) \tau_y.
\end{align}

We first investigate the surface $\textrm{\Rmnum{1}}$. By expressing $k_y$ as $-i \partial_{y}$, we can rewrite the Hamiltonian $ \bar{H}_\textrm{cm}(\mathbf{k}) $ as $ \bar{H}_\textrm{cm}(k_x, -i\partial_y, k_z) = \tilde{H}_{\textrm{cm},1}(k_x, -i\partial_y, k_z) + \tilde{H}_{\textrm{cm},2}(k_x, -i\partial_y, k_z) $, where
\begin{align}\label{EffHEq2}
\tilde{H}_{\textrm{cm},1}(k_x, -i\partial_y, k_z)  =  \left(m + 3 t +\frac{t}{2} \partial_{y}^2 \right) \tau_z - i \Delta_1 \partial_{y} \sigma_y \tau_x
\end{align}
is Hermitian, and
\begin{align}
\tilde{H}_{\textrm{cm},2}(k_x, -i\partial_y, k_z)  =  \Delta_1 \left( k_x \sigma_x + k_z \sigma_z \right) \tau_x + i \gamma_0  \left(\sigma_x + \sigma_y \right) \tau_x + i \gamma_z \sigma_z \tau_x- \frac{\Delta_2}{2} \tau_y \partial_{y}^2  
\end{align}
is treated as a perturbation for $\max \{\Delta_2, \abs{\gamma_0}, \abs{\gamma_z}\} \ll \max \{\abs{m + 3t}, t/2, \Delta_1 \} $. Note that we have neglected insignificant terms $k_x^2$ and $k_z^2$ for $\mathbf{k}$ around the $\Gamma$ point.
To obtain the eigenvalue equation $\tilde{H}_{\textrm{cm},1} \phi_{\textrm{\Rmnum{1}}}(y) = E \phi_{\textrm{\Rmnum{1}}}(y)$, with $E = 0$ subject to the boundary condition $\phi_{\textrm{\Rmnum{1}}}(0) = \phi_{\textrm{\Rmnum{1}}}(+\infty) =0$, we write the solution in the following form:
\begin{align}
\phi_{\textrm{\Rmnum{1}}}(y) = \mathcal{N}_y\sin(\alpha y) \textrm{e}^{-\beta y} \textrm{e}^{i \left(k_x x + k_z z\right)} \chi_{\textrm{\Rmnum{1}}}, ~~ \text{Re}(\beta) > 0,
\end{align}
where the normalization constant is given by $\mathcal{N}_y = 2 \sqrt{\beta (\alpha^2 + \beta^2)/\alpha^2}$, and the eigenvector $\chi_{\textrm{\Rmnum{1}}}$ satisfies $\sigma_y \tau_y \chi_{\textrm{\Rmnum{1}}} =  \chi_{\textrm{\Rmnum{1}}}$.  Then the effective Hamiltonian for the surface $\textrm{\Rmnum{1}}$ can be obtained in this basis as
\begin{align}
\mathcal{H}_{\textrm{surf}}^{\textrm{\Rmnum{1}}}  = \int_{0}^{+\infty} \!\! \phi_{\textrm{\Rmnum{1}}}^*(y) \tilde{H}_{\textrm{cm},2} \phi_{\textrm{\Rmnum{1}}}(y) \: dy .
\end{align}
Therefore, we have
\begin{align}\label{SurfaceEq3}
\mathcal{H}_{\textrm{surf}}^{\textrm{\Rmnum{1}}} = \Delta_2 \left(3 + \frac{m}{t}\right) \varrho_z - \left(\Delta_1 k_x + i\gamma_0 \right) \varrho_x - \left(\Delta_1 k_z + i \gamma_z \right) \varrho_y.
\end{align}

The low-energy effective Hamiltonians for the surfaces $\textrm{\Rmnum{2}}$, $\textrm{\Rmnum{3}}$ and $\textrm{\Rmnum{4}}$ can be obtained by the same procedures 
\begin{align}\label{SurfaceEq4}
\mathcal{H}_{\textrm{surf}}^{\textrm{\Rmnum{2}}}  = \Delta_2 \left(3 + \frac{m}{t}\right) \varrho_z + \left(\Delta_1 k_y + i\gamma_0 \right) \varrho_x + \left(\Delta_1 k_z + i \gamma_z \right) \varrho_y, 
\end{align}
\begin{align}\label{SurfaceEq5}
\mathcal{H}_{\textrm{surf}}^{\textrm{\Rmnum{3}}}  = \Delta_2 \left(3 + \frac{m}{t}\right) \varrho_z + \left(\Delta_1 k_x + i\gamma_0 \right) \varrho_x - \left(\Delta_1 k_z + i \gamma_z \right) \varrho_y, 
\end{align}
\begin{align}\label{SurfaceEq6}
\mathcal{H}_{\textrm{surf}}^{\textrm{\Rmnum{4}}}  = \Delta_2 \left(3 + \frac{m}{t}\right) \varrho_z - \left(\Delta_1 k_y + i\gamma_0 \right) \varrho_x + \left(\Delta_1 k_z + i \gamma_z \right) \varrho_y,
\end{align}
for $ 3 + m/t > 0 $.

For the Hermitian case (i.e., $\gamma_0 = 0$), the last two kinetic terms of the effective Hamiltonian in Eqs.~(\ref{SurfaceEq3})-(\ref{SurfaceEq6}) describe the gapless surface states, which are gapped out by the first terms with the Dirac mass. To derive the second-order boundary modes, we introduce a new coordinate $p$ along anticlockwise direction within the planes parallel to the $xy$ plane, and we rewrite Eqs.~(\ref{SurfaceEq3})-(\ref{SurfaceEq6}) as
\begin{align}\label{SurfaceEq7}
\mathcal{H}_{\textrm{surf}}  = \Delta_2 \left(3 + \frac{m}{t}\right) \varrho_z - \left[- i \Delta(p) \partial p + i \gamma(p) \right] \varrho_x - \left(\Delta_1 k_z + i \gamma_z \right) \varrho_y,
\end{align}
where $\Delta(p) = \Delta_1, -\Delta_1, \Delta_1 , -\Delta_1$, and $\gamma(p) = \gamma_0, -\gamma_0, -\gamma_0, \gamma_0$ along the anticlockwise direction of the four surfaces.  According to Eq.~(\ref{SurfaceEq7}), it is easy to verify that the Hermitian part (i.e., $\gamma_0 = 0$) of $H_\textrm{3D}(\mathbf{k})$ supports four-fold degenerate gapless hinge modes for $k_z = 0$ (analogous to the Jackiw-Rebbi model \cite{SMPhysRevD.13.3398}). 

In the presence of non-Hermiticity, we first solve the boundary mode along the hinge intersected by the surfaces $\textrm{\Rmnum{1}}$ and $\textrm{\Rmnum{2}}$. We write this boundary mode in the following form:
\begin{align}\label{SurfaceEq8}
\Psi_{h}^{\textrm{\Rmnum{1}}} = \textrm{e}^{i k_z z + \kappa_{\textrm{\Rmnum{1}},1} \tilde{p}} \chi_{h,1}, ~~~~~ \textrm{Re}(\kappa_{\textrm{\Rmnum{1}},1}) > 0,  ~~~~~\tilde{p} = p - L < 0,  
\end{align}
\begin{align}\label{SurfaceEq9}
\Psi_{h}^{\textrm{\Rmnum{2}}} = \textrm{e}^{i k_z z + \kappa_{\textrm{\Rmnum{2}},1} \tilde{p}} \chi_{h,1}, ~~~~~ \textrm{Re}(\kappa_{\textrm{\Rmnum{2}},1}) < 0,  ~~~~~ \tilde{p} = p - L > 0,  
\end{align}
where $L$ is the width of cubic sample, and $\chi_{h,1}$ is the eigenvector. To find the gapless boundary mode along the hinge intersected by the surfaces $\textrm{\Rmnum{1}}$ and $\textrm{\Rmnum{2}}$, the following equations are satisfied:
\begin{align}\label{SurfaceEq10}
\kappa_{\textrm{\Rmnum{1}},1} = \frac{ \gamma_0 \pm \sqrt{\delta^2 + \left(\Delta_1 k_z + i \gamma_z \right)^2}}{\Delta_1}, ~~~~~  \kappa_{\textrm{\Rmnum{2}},1} = \frac{\gamma_0 \pm \sqrt{\delta^2 + \left(\Delta_1 k_z + i \gamma_z \right)^2}}{\Delta_1}, ~~~~~ \textrm{Re}(\kappa_{\textrm{\Rmnum{1}},1}) > 0, ~~~~~ \textrm{Re}(\kappa_{\textrm{\Rmnum{2}},1}) < 0,
\end{align}
where $\delta = \Delta_2 \left(3 + m/t \right)$.

By applying the same procedures, to have the zero gapless boundary modes along hinges intersected by surfaces $\textrm{\Rmnum{2}}$ and $\textrm{\Rmnum{3}}$, surfaces $\textrm{\Rmnum{3}}$ and $\textrm{\Rmnum{4}}$ and surfaces $\textrm{\Rmnum{4}}$ and $\textrm{\Rmnum{1}}$, we have the following equalities: 
\begin{align}\label{SurfaceEq11}
\kappa_{\textrm{\Rmnum{2}},2} = \frac{ \gamma_0 \pm \sqrt{\delta^2 + \left(\Delta_1 k_z + i \gamma_z \right)^2}}{\Delta_1}, ~~~~~  \kappa_{\textrm{\Rmnum{3}},2} = \frac{-\gamma_0 \pm \sqrt{\delta^2 + \left(\Delta_1 k_z + i \gamma_z \right)^2}}{\Delta_1}, ~~~~~ \textrm{Re}(\kappa_{\textrm{\Rmnum{2}},2}) > 0, ~~~~~ \textrm{Re}(\kappa_{\textrm{\Rmnum{3}},2}) < 0,
\end{align}
\begin{align}\label{SurfaceEq12}
\kappa_{\textrm{\Rmnum{3}},3} = \frac{ -\gamma_0 \pm \sqrt{\delta^2 + \left(\Delta_1 k_z + i \gamma_z \right)^2}}{\Delta_1}, ~~~~~  \kappa_{\textrm{\Rmnum{4}},3} = \frac{-\gamma_0 \pm \sqrt{\delta^2 + \left(\Delta_1 k_z + i \gamma_z \right)^2}}{\Delta_1}, ~~~~~ \textrm{Re}(\kappa_{\textrm{\Rmnum{3}},3}) > 0, ~~~~~ \textrm{Re}(\kappa_{\textrm{\Rmnum{4}},3}) < 0,
\end{align}
\begin{align}\label{SurfaceEq13}
\kappa_{\textrm{\Rmnum{4}},4} = \frac{ -\gamma_0 \pm \sqrt{\delta^2 + \left(\Delta_1 k_z + i \gamma_z \right)^2}}{\Delta_1}, ~~~~~  \kappa_{\textrm{\Rmnum{1}},4} = \frac{\gamma_0 \pm \sqrt{\delta^2 + \left(\Delta_1 k_z + i \gamma_z \right)^2}}{\Delta_1}, ~~~~~ \textrm{Re}(\kappa_{\textrm{\Rmnum{4}},4}) > 0, ~~~~~ \textrm{Re}(\kappa_{\textrm{\Rmnum{1}},4}) < 0.
\end{align}
%

According to Eqs.~(\ref{SurfaceEq10})-(\ref{SurfaceEq13}), it is straightforward to verify that the gapless boundary modes, under the periodic boundary condition along the $z$ direction, appear only along the hinge intersected by the surfaces $\textrm{\Rmnum{2}}$ and $\textrm{\Rmnum{3}}$  if $\abs{\gamma_0} > \sqrt{\delta^2 - \gamma_z^2}$ with $\abs{\delta} > \abs{\gamma_z}$ or $\abs{\gamma_z} > \abs{\delta}$ . In summary, the simple low-energy effective Hamiltonian presented here can explain the existence and the localization at one hinge of second-order gapless boundary modes due to the interplay between mirror-ration symmetry and non-Hermiticity.

\subsection{Topological index} 
As shown in the main text, due to mirror-rotation symmetry $\mathcal{M}_{xy}$, we can write the Hamiltonian $H_\textrm{3D}(\mathbf{k})$ into the block-diagonal form with $k_x = k_y = k$ as

\begin{align}\label{wnEq3DTIEq1}
\tilde{U}^{-1} H_\textrm{3D}(\mathbf{k}) \tilde{U} = \left[\begin{matrix}
H_{+}(k,k_z) & 0 \\
0 & H_{-}(k,k_z) \\  \end{matrix}\right],
\end{align}
where $H_{+}(k,k_z)$ acts on the $+i$ mirror-rotation subspace, and $H_{-}(k,k_z)$ acts on the $-i$ mirror-rotation subspace. The unitary transformation $ \tilde{U} $ is
\begin{align}
\tilde{U} = \left[\begin{matrix}
0 & -\frac{1+i}{2} & 0 & \frac{1+i}{2} \\
\frac{1+i}{2} & 0 & -\frac{1+i}{2} & 0 \\
0 & \frac{1}{\sqrt{2}} & 0 & \frac{1}{\sqrt{2}} \\
\frac{1}{\sqrt{2}} & 0 & \frac{1}{\sqrt{2}} & 0  \\  \end{matrix}\right],
\end{align}
and $H_{+}(k,k_z)$ and $H_{-}(k,k_z)$ are
\begin{align}\label{wnEq3DTIEq2}
H_{+}(k,k_z) = -\left[m + 2 t \cos(k) + t \cos(k_z)\right] \sigma_z + \sqrt{2}\left[\Delta_1 \sin(k) + i  \gamma_0 \right]  \sigma_y - \left[\Delta_1 \sin(k_z) + i \gamma_z \right]  \sigma_x,
\end{align}
\begin{align}\label{wnEq3DTIEq3}
H_{-}(k,k_z) = -\left[m + 2 t \cos(k) + t \cos(k_z)\right] \sigma_z -  \sqrt{2}\left[\Delta_1 \sin(k) + i  \gamma_0 \right]  \sigma_y - \left[\Delta_1 \sin(k_z) + i \gamma_z \right] \sigma_x. 
\end{align}

To account for the non-Bloch-wave behavior of the non-Hermitian systems and to capture the essential physics with analytical results, we consider a low-energy continuum mode of $H_\textrm{3D}(\mathbf{k})$ and $H_{\pm}(k,k_z)$. Then the non-Bloch-wave behavior can be taken into account by replacing the real vavevector $k$ in Eqs.~(\ref{wnEq3DTIEq1})-(\ref{wnEq3DTIEq3}) with the complex one after applying the degenerate perturbation theory to Hamiltonian $H_\textrm{3D}$ with open boundaries along the $x$, $y$ and $z$ directions for $\min\{\abs{m+3t},\, \abs{t}, \, \abs{\Delta_1} \} \gg \max \{\abs{\gamma_0}, \abs{\gamma_z}\}$ \cite{SMYaoarXiv:1804.04672}:
\begin{align}\label{wnEq3DTIEq210}
k \, \to \,  k + i k', ~~~~  k_z \, \to \,  k_z + i k'_z,
\end{align}
with 
\begin{align}\label{wnEq3DTIEq211}
k' = - \frac{\gamma_0}{\Delta_1}, ~~~~~~~ k'_z = - \frac{\gamma_z}{\Delta_1}.~
\end{align}
Then, we rewrite Eq.~(\ref{wnEq3DTIEq2}) and (\ref{wnEq3DTIEq3}) as
\begin{align}\label{wnEq3DTIEq212}
\bar{H}_{+}(k,k_z) = -\left[m + 3 t - t \left(k - i \frac{\gamma_0}{\Delta_1}\right)^2- \frac{t}{2}k_z^2 \right] \sigma_z + \sqrt{2}\left[\Delta_1 \left(k - i \frac{\gamma_0}{\Delta_1}\right) + i  \gamma_0 \right]  \sigma_y - \left[\Delta_1 \left(k_z- i \frac{\gamma_z}{\Delta_1}\right) + i \gamma_z \right] \sigma_x,
\end{align}
\begin{align}\label{wnEq3DTIEq312}
\bar{H}_{-}(k,k_z) = -\left[m + 3 t - t \left(k - i \frac{\gamma_0}{\Delta_1}\right)^2- \frac{t}{2} k_z^2 \right] \sigma_z -  \sqrt{2}\left[\Delta_1 \left(k - i \frac{\gamma_0}{\Delta_1}\right) + i  \gamma_0 \right]  \sigma_y - \left[\Delta_1 \left(k_z- i \frac{\gamma_z}{\Delta_1}\right) + i \gamma_z \right] \sigma_x. 
\end{align}

The Chern numbers $C_{+}$ and $C_{-}$, corresponding to the $\bar{H}_{+}(k,k_z)$ and $\bar{H}_{-}(k,k_z)$, are defined as
\begin{align}\label{wnEq3DTIEq4}
C_{\pm} = \frac{1}{2\pi} \int_{\textrm{BZ}} \!\! F_{\pm}(k, k_z) ~dkdk_z,
\end{align}
where $F_{\pm}(k, k_z)$ is the Berry curvature
\begin{align}\label{wnEq3DTIEq5}
F_{\pm}(k, k_z) = {\partial_k} A_{\pm}^{k_z}(k, k_z) - {\partial_{k_z}} A_{\pm}^{k}(k, k_z),
\end{align}
and $A_{\pm}^{\mu}$ is the Berry connection
\begin{align}\label{wnEq3DTIEq6}
A_{\pm}^{\mu}(k, k_z) = i \bra{\chi_{\pm}(k, k_z)}  \ket{\partial_{\mu} \phi_{\pm}(k, k_z)},
\end{align}
where $\ket{\phi_{\pm}^{\alpha}(\mathbf{k})}$ and $\ket{\chi_{\pm}^{\alpha}(\mathbf{k})}$ are the right and left eigenstates of $\bar{H}_{\pm}(k, k_z)$. Then, the total Chern number is given by
\begin{align}\label{wnEq3DTIEq7}
C = C_{+} - C_{-}.
\end{align}
%


%

\end{document}